\documentclass{aa}  
\usepackage{xcolor}
\usepackage[colorlinks=true, allcolors=cyan]{hyperref}
\usepackage{subcaption}
\usepackage{lscape}
\usepackage{graphicx}
\usepackage{txfonts}
\begin{document}

   \title{Starburst galaxies in the Hydra I cluster}

  \author{
  Clara C. de la Casa\inst{\ref{iaa},\thanks{Corresponding author: clarac@iaa.es}}, Kelley M.~Hess\inst{\ref{chalmers},\ref{astron},\ref{iaa}}
  , Lourdes Verdes-Montenegro\inst{\ref{iaa}},
  Ralf Kotulla\inst{\ref{uwisc}},
  Hao Chen\inst{\ref{zjlab},\ref{uct}},
  Tom H. Jarrett\inst{\ref{haw},\ref{uct},\ref{wsu}},
  Michelle E. Cluver\inst{\ref{cas_aus},\ref{pa_sa}},
  Simon B. De Daniloff\inst{\ref{ugr},\ref{iram}},
  Marie-Lou Gendron-Marsolais\inst{\ref{iaa},\ref{laval}},
  Claude Carignan\inst{\ref{uct},\ref{mont},\ref{ouag}},
  John S. Gallagher III\inst{\ref{uwisc}},
  Renée C. Kraan-Korteweg\inst{\ref{uct}},
  \and
  Roger Ianjamasimanana \inst{\ref{iaa}}}
    \authorrunning{de la Casa, C.~C.~et al.}
    
    \institute{Instituto de Astrof\'{i}sica de Andaluc\'{i}a (CSIC), Glorieta de la Astronom\'{i}a s/n, 18008 Granada, Spain\label{iaa}
    \and Department of Space, Earth and Environment, Chalmers University of Technology, Onsala Space Observatory, 43992 Onsala, Sweden \label{chalmers}
    \and ASTRON, the Netherlands Institute for Radio Astronomy, Postbus 2, 7990 AA, Dwingeloo, The Netherlands\label{astron}
    \and Department of Astronomy, University of Wisconsin-Madison, 475 N Charter St, Madison, WI, 53706, USA\label{uwisc}
    \and Research Center for Astronomical Computing, Zhejiang Laboratory, Hangzhou 311100, China\label{zjlab}
    \and Department of Astronomy, University of Cape Town, Private Bag X3, 7701 Rondebosch, South Africa\label{uct}
    \and Institute for Astronomy, University of Hawaii at Hilo, 640 N Aohoku Pl 209, Hilo, HI 96720, USA\label{haw}
    \and Western Sydney University, Locked Bag 1797, Penrith South DC, NSW 1797, Australia\label{wsu}
    \and Centre for Astrophysics and Supercomputing, Swinburne University of Technology,John Street, Hawthorn, 3122, Australia\label{cas_aus}
    \and Department of Physics and Astronomy, University of the Western Cape, Robert Sobukwe Road, Bellville, 7535, South Africa\label{pa_sa}
    \and Dpto. de Física Teórica y del Cosmos, University of Granada, Facultad de Ciencias (Edificio Mecenas), 18071 Granada, Spain \label{ugr}
    \and Institut de Radioastonomie Millimétrique (IRAM), Av. Divina Pastora 7, Núcleo Central 18012, Granada, Spain \label{iram}
    \and D\'{e}partement de physique, Universit\'{e} de Montr\'{e}al,  Complexe des sciences MIL, 1375 Avenue Th\'{e}r\`{e}se-Lavoie-Roux, Montr\'{e}al, Qc, Canada H2V 0B3 \label{mont}
    \and Laboratoire de Physique et de Chimie de l'Environnement, Observatoire d'Astrophysique de l'Universit\'{e}  Ouaga I Pr Joseph Ki-Zerbo (ODAUO), BP 7021, Ouaga 03, Burkina Faso\label{ouag}
    \and Département de physique, de génie physique et d’optique, Université Laval, Québec (QC), G1V 0A6, Canada \label{laval}
         \and 
         Inter-University Institute for Data Intensive Astronomy (IDIA), University of Cape Town, Rondebosch, Cape Town, 7701, South Africa\label{idia}
         }

   \date{Received ---; accepted ---}

 
  \abstract
  {Studying the impact of environment on star formation and quenching pathways requires statistically relevant samples of galaxies in a wide mass range. We present a new catalog of 196 galaxies of the nearby Hydra I cluster out to $\sim$1.75$\rm r_{200}$, consisting of broad u,g,r,i,z along with narrowband H${\alpha}$ measurements. These deep optical images were obtained with the DECam camera (CTIO) and reach down to a surface brightness limit of $\mu( 3\sigma;10''\times10'')$=26.9 mag $\rm arcsec^2$ in the g band. We also report the HI properties for 89 cluster members detected with MeerKAT. A color magnitude diagram (CMD) shows a bimodal distribution typical of a cluster population, more evolved than those found in isolation. We combine optical H${\alpha}$ and WISE infrared data to compare the star formation history at two distinct timescales. Differences in the star forming activity depicted by both populations manifest as starburst in 24 found members. Of these, 18 starburst galaxies have neutral gas measurements, and show disturbed HI disks that suggest an environmentally-triggered boost in star formation within the last 10$^7$ yrs. Processes such as ram pressure stripping or tidal interactions may underlie their enhanced star-forming activity and asymmetric disks. Since Hydra's dynamical history is unclear, we examine the spatial and velocity distribution of the sample. We reveal a possible link between the large scale structure feeding the Hydra I cluster and the heightened star-forming activity of the starburst galaxies. This feeding pattern matches the few substructure that has been identified in Hydra in previous works, and may explain its origin. Our results portray a picture of a cluster with an evolved nature, plus a population of new infalling galaxies that manifest the impact of their first contact with the cluster environment through star formation, color, morphology and gas content transformations. }


   \maketitle
%

\section{Introduction}

Star formation has traditionally dominated the study of galaxy evolution as one of the most reliable indicators of the processes that fuel a galaxy's growth \citep{Kouroumpatzakis2021}. For galaxies in isolation, the analysis of the mechanisms that regulate the conversion of interstellar medium (ISM) into stars provides a clear window into a secular evolution governed by factors such as gas density or metallicity, and processes like feedback mechanisms associated to star formation and supernovae \citep{athanassoula2006barred,kormendy2008secular}. However, since the morphology-density relation was first observed, it became clear that the environment plays a major role in shaping the morphological and chemical evolution of non-isolated galaxies over cosmic time \citep{dressler1980galaxy,postman1984morphology,springel2005simulations}. Unveiling the specifics behind the transformations caused by gravitational and hydrodynamical interactions with the environment  remains as a major challenge in the field of galaxy evolution.

The most extreme, large-scale structures to probe the effect of the environment upon galaxies are galaxy clusters \citep{dressler1980galaxy,postman1984morphology,springel2005simulations}. These systems of gas, dark matter and galaxies continue to accrete material from the cosmic web through extensive filaments to the present day \citep{castignani2022virgo,hyeonghan2024weak}.\\ 
The analysis of the impact of the interaction with other galaxies and/or the intracluster medium \citep{Paulino-Afonso2019} on star formation has revealed how galaxies age prematurely when exposed to some of the most extreme environmental processes in the cosmos. Galaxy-galaxy interactions include processes such as harassment \citep{spitzer1951stellar,richstone1976collisions,farouki1981computer,moore1996galaxy, moore1998morphological}, cannibalism \citep{ostriker1975another,hausman1978galactic,malumuth1984evolution}, tidal stripping \citep{merritt1983relaxation, merritt1984relaxation,malumuth1984evolution} or merging \citep{negroponte1983simulations, barnes1991fueling, barnes1992dynamics,barnes1996transformations, mihos1994ultraluminous,hernquist1995excitation}. Meanwhile, galaxy and intracluster medium (ICM) interactions involve mechanisms like thermal evaporation \citep{cowie1977thermal, nipoti2007role, nipoti2009cool, nipoti2010cusp}, starvation \citep{larson1980evolution, balogh2000origin} or ram pressure stripping \citep{gunn1972infall, nulsen1982transport,Quilis2000}. Galaxies undergoing such processes manifest physical and structural perturbations: removal of angular momentum \citep{sorgho2024amiga}, asymmetric disks and/or tails of ripped material \citep{serra2024meerkat}, redder colors, altered stellar masses and star formation rates \citep{Lima-Dias2024}. In general, all these processes combine to increase the removal of cold and warm gas from galaxies, strongly suppressing star formation in a process also known as quenching. \\
A common reference frame to identify the impact of external mechanisms and trace the evolutionary history of galaxies comes from  studies that simulate \citep{stinson2006star,torrey2020impact,walters2021structural} and catalog \citep{haynes1984neutral, vavilova2009morphological, hernandez2010unam, argudo2015catalogues} the slow evolution driven by internal mechanisms for galaxies in isolation.
Indeed, it is widely accepted that field galaxies are more star-forming than those in the core of clusters, which are usually passively evolving \citep{lewis20022df}. \cite{dressler1980galaxy} showed that environmental density was proportional to the fraction of early type galaxies (ETGs) and inversely proportional to the fraction of late type galaxies (LTGs). This also manifests as a bimodality in the distribution of galaxy populations in clusters: a red sequence of galaxies in the far end of the Hubble sequence, and a blue cloud of galaxies that are newcomers or in their first passage after going through the cluster. The spectrum between the red sequence and blue cloud, known as green valley, is usually undersampled. This prevents us from understanding how different populations evolve across the Hubble sequence and what is the exact role of the environment at each stage \citep{schawinski2014green}.

 Consequently, it is imperative to conduct statistically relevant studies capable of offering insight on how environment promotes quenching as a function of stellar mass, morphology, distance to the cluster core, local density, and time since first infall.

 Simulations have already shown how the mass of a cluster is directly proportional to the velocity at which galaxies start to quench after first infall \citep{pallero19}. They have also been able to recreate how pre-processing in filaments limits the stellar production in galaxies prior to entering the cluster, and contribute to increase the proportion of members of the red sequence in these systems \citep{salerno2022star,donnari2021quenched}. This effect is particularly relevant for satellite galaxies, whereas more central galaxies suffer from more dramatic quenching in situ. In general, these simulations agree with observations, that have also shown how star formation rates are suppressed in filaments when compared to less dense environments \citep{Martinez16SSDS}. All these processes result in galaxies entering the cluster at a variety of evolutionary stages. This increases the challenge of separating pre-processing and cluster-driven mechanisms. The latter have been traditionally characterized in a variety of clusters from the point of view of star formation quenching \citep{cybulski2014voids, donnari2021quenched} as well as morphology \citep{su2021fornax, oh2018kydisc}, and cold gas deficiencies and dynamics \citep{upadhyay2021star, kleiner2021meerkat,cortese2021dawes}. 

 The Hydra cluster (Abell 1060) is one of such clusters. As the fifth nearest cluster in the local universe, it belongs to the Great Attractor superstructure, where it shares a filamentary connection with ate least three well-characterized clusters: Antlia \citep[z = 0.0087;][]{Hess2015}, Norma \citep[z = 0.01570;][]{Courtois2012}, and Centaurus \citep[z = 0.01570;][]{courtois2013cosmography}.
 
Hydra's dynamical history is unclear. Authors as \cite{fitchett1988dynamics} concluded that Hydra I has not recently suffered major mergers and that, in general, it is perturbed but close to relaxation. Other studies observed that Hydra I is a relatively gas-rich cluster \citep{mcmahon1992hi} with ongoing processes of infall of at least 2 substructures \citep{LaMarca2022}, and merging processes close to the galactic core as shown by the presence of a remarkable number of dwarfs in \citep{LaMarca2022} and planetary nebulae in \citep{ventimiglia2011unmixed, arnaboldi2012tale}. Hence, Hydra I has the potential to provide rich insight into the wide variety of phenomena derived from the interplay between galaxies and their environment, particularly when combining observations at different wavelengths. Studies focused on optical emission such as \citep{Lima-Dias2019} found that almost 90\% of Hydra I galaxies are quenched, while only a rough 23\% from the total were late-type galaxies. These authors made the point that physical properties, such as star formation, change faster than morphological properties in clusters. In \citep{Lima-Dias2024}, ram pressure stripping and tidal interactions were shown to be responsible of the interruption of star formation after performing a bulge-disk decomposition to a sample of galaxies undergoing compaction. On the other hand, works combining measurements from several wavelengths were capable of providing additional information relating the effects of environment on the gas reservoirs of young galaxies and their star formation.\cite{Wang2021} employed WALLABY HI measurements in combination with optical Pan-STARRS bands to find that more than 70\% of Hydra's galaxies with detectable levels of neutral gas within a virial radius are submitted to at least weak levels of ram pressure stripping, although the effects of such mild interactions are not certain. Likewise, \cite{Reynolds2021} examined the Hydra I cluster in combination with optical Pan-STARRS and infrared WISE data, and concluded that the environmental impact on the neutral gas reservoir can start as far as 1.5$r_{\rm 200}$ from the cluster core, affecting first the external disks of galaxies, where a reduction in recent star formation rates was found even for main sequence galaxies. Later, \cite{Hess2022} combined the analysis of star formation rates from the H$\alpha$ optical narrowband with resolved measurements of the neutral gas content, and found conclusive evidence of ram pressure acting upon a group of recently accreted galaxies near the cluster core, and even active star formation in the stripped material that could suggest an outcome for the lost gas.

In this work, we expand the previous analysis and combine optical and infrared tracers of star formation to evaluate the impact of environment on galaxies at different timescales out to $\sim$1.75 virial radius. We employ commissioned optical broadband and H$\alpha$ narrowband images to trace star formation in timescales of 10$^{7}$ yr, and archival infrared data to compute star formation rates in timescales of 10$^{9}$ yr. Our aim is to show the contrast between these 2 sets of measurements (optical and infrared) and a control sample of isolated galaxies. Additionally, we present new radio neutral atomic hydrogen (HI) measurements of the Hydra cluster. The cold gas content and disk morphology of these galaxies are used to link their observed star formation activities to their interactions with the environment. Throughout this multiwavelength study, we will consider a $\Lambda$CMD cosmology with $\Omega_{\rm m}= 0.27$, $\Omega_{\Lambda} = 0.73$ and $H_0 = 70$ km s$^{-1}$ $Mpc^{-1}$. We will assume a redshift of $z_{\rm HC}\approx0.012$ for the Hydra I cluster, corresponding to a distance of $d_{\rm HC}=58.6$ Mpc \citep{Tully2015}, and a characteristic radius of $r_{200}=1.35$ Mpc containing a mass of $M_{\rm 200}=3.02\times10^{14}M_{\odot}$ \citep{reiprich2002mass}. The velocity dispersion of the cluster is $\sigma_{\rm HC}=620$ km s$^{-1}$ \citep{Wang2021}. Throughout this work, we will assume that the cluster core is at the location of NGC 3311, the brightest cluster galaxy: $\rm RA_{core}=159.17^{\circ}$, $\rm DEC_{core}=-27.52^{\circ}$.

The structure of this work is as follows. In Section \ref{sec:Data} we provide a description of the data and the selection of the sample that will be analyzed throughout the paper. In Section \ref{sec:SF}, we separate the sample in different populations based on their star formation activity. This serves to identify galaxies with altered levels of star formation, which may be a manifestation of their interaction with the environment. In section \ref{sec:discuss}, we discuss our classification. In section \ref{subsec:sfprojdist} we study the dependence of star formation with projected distance to the center of the cluster. Section \ref{subsec: asymmetry} provides an analysis of a 2D spatial distribution of the galaxies in the cluster. In section \ref{subsec:substructure} we show possible substructures inside Hydra I, and discuss their link to larger scale structure feeding the cluster and how it might affect the star formation rates of our sample. We have studied the morphology and color of galaxies in section \ref{subsec:morphcol}, and their velocity distribution in section \ref{subsec:ppsd}. We analyze how the cold gas disk morphologies of galaxies with heightened star formation rates can be related to environmental process in section \ref{subsec:himorph}. Finally, in section \ref{sec:summary} we summarize our observations and present the main conclusions of this work.

\section{Data}\label{sec:Data}

This study is based on a redshift-selected sample of galaxies. The data comes from a combination of all-sky public surveys (6dFGS, WISE), and targeted observations (DECam, MeerKAT). Our parent sample consists of 282 galaxies with a velocity in the range $v\in\left[2000,6000\right]km/s$, roughly  $\pm3\sigma_{\rm HC}$.

\begin{figure}[ht!]
    \centering
    \includegraphics[width=\columnwidth]{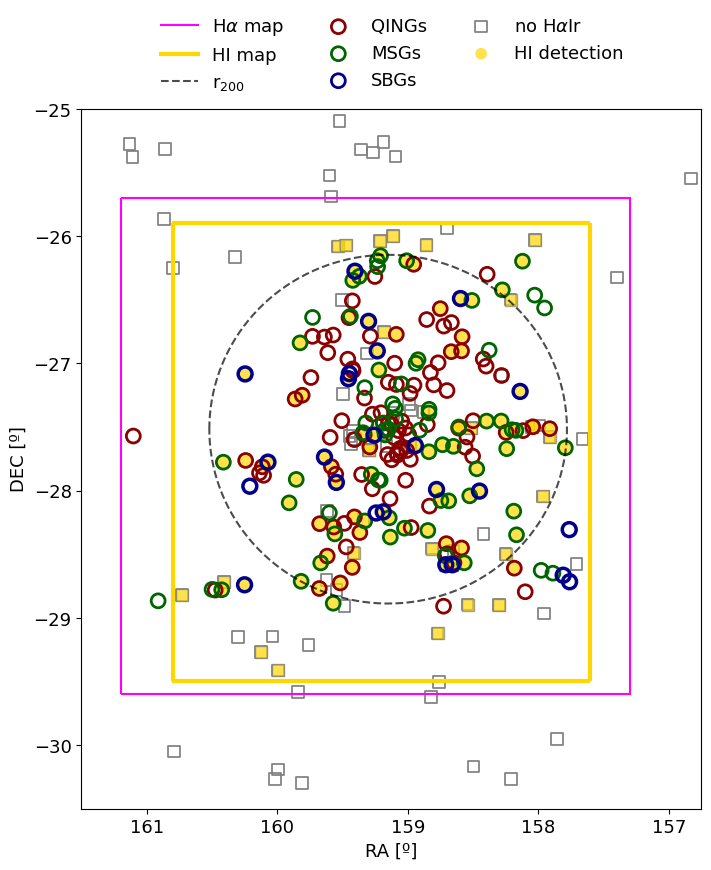}
    \caption{Galaxies from our parent sample in equatorial coordinates. The black dashed circle is the virial radius $r_{\rm 200}$. The fuchsia and yellow squared regions show the coverage of our DECam and HI measurements, respectively.\\  Gray squares represent the galaxies that do not belong to the H$\alpha$IR sample, but are members of Hydra I according to either 6dFGS or NED (see sec. \ref{sec:Data}). Colorized markers are discussed in sections \ref{sec:SF} and \ref{subsec:ppsd}, where we separate our sample in three populations according to their star formation activity relative to the main sequence (see section \ref{sec:DMS}): starburst galaxies (SBGs), main sequence galaxies (MSGs), and quenching galaxies (QINGs).
    Yellow filled markers show those galaxies that were detected in HI. \\
    }
    \label{fig:map}
\end{figure}

\begin{figure}
    \centering
    \includegraphics[width=\columnwidth]{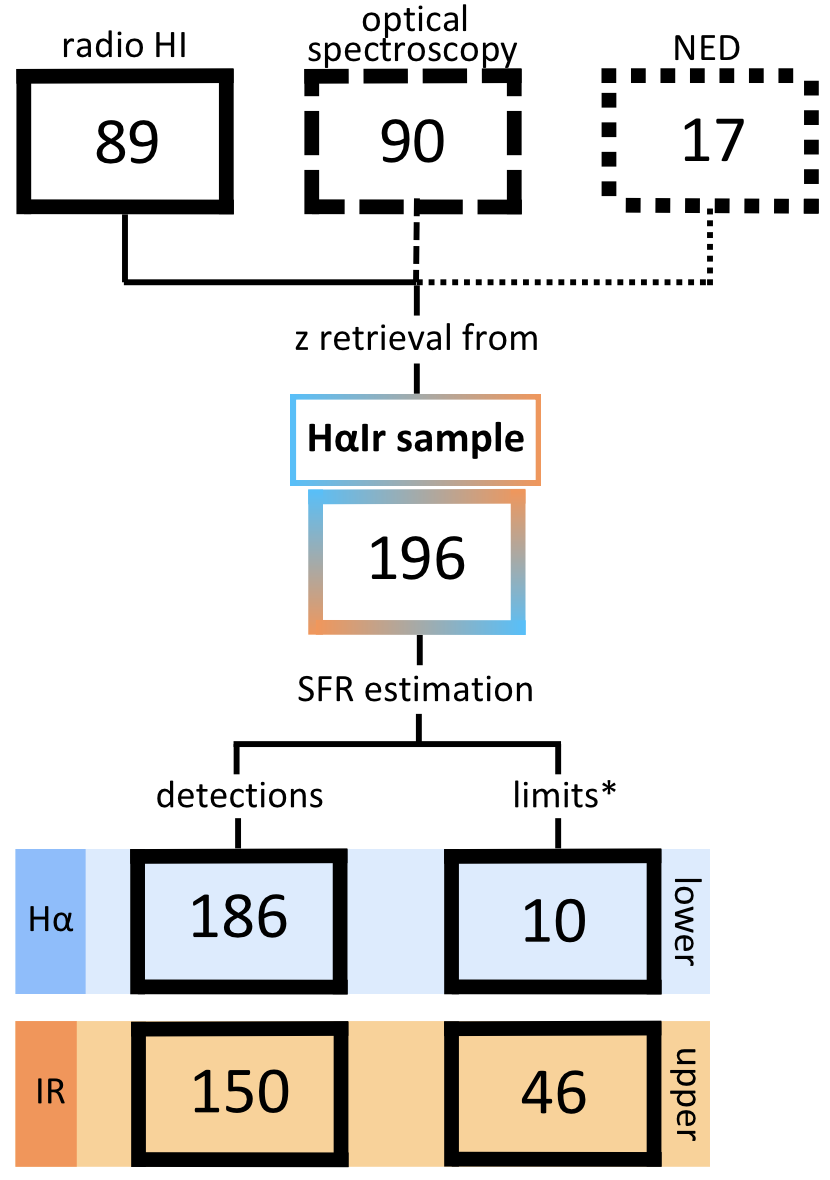}
    \caption{Diagram of the selection process for the 196 Hydra cluster members from the final sample. \\
    Center-up: The selection of the 196 galaxies for this study was based on both optical and infrared data availability. Redshifts come, in order of preference: from radio HI estimations for the 89 galaxies with an HI detection, from 6dFGS for another 90 galaxies without an HI detection, and finally from NED for 17 galaxies without HI and spectroscopic detections.\\
    Center-down: Star formation rates (SFR) were calculated from 186 H$\alpha$ and 150 mIR confident detections. (*) 10 estimations of the SFR$_{\rm H\alpha}$ are lower limits due to high background signal from nearby galaxies, lower S/N, uncertain continuum subtraction, or the presence of very bright nearby sources. We include 46 estimations of SFR$_{\rm IR}$ based on upper limits measurements due to low S/N.}
    \label{fig:tabla1}
\end{figure}

The different samples for each wavelength and their particular extent are shown in Fig. \ref{fig:map}. They will be discussed in depth, as well as the process of selection of the final 196 galaxies, in the following subsections (see Fig. \ref{fig:tabla1} for an outline).
MeerKAT provides (high quality) redshifts for 114 galaxies in the parent sample. For the remaining galaxies, 137 redshifts were retrieved from optical spectroscopy of 6dFGS, and 29 from NED database. The morphologies of the galaxies were also extracted from NED. Additional redshifts are known for 20 ultra-diffuse galaxies (UDGs) from the LEWIS ESO program, that were recently reported in \citep{iodice2023looking}. We do not include them in this study as their low emission in the H$\alpha$ narrowband prevents an accurate estimation of their star-forming activity.


 \begin{figure*}[ht!]
    \centering
    \includegraphics[width=18cm]{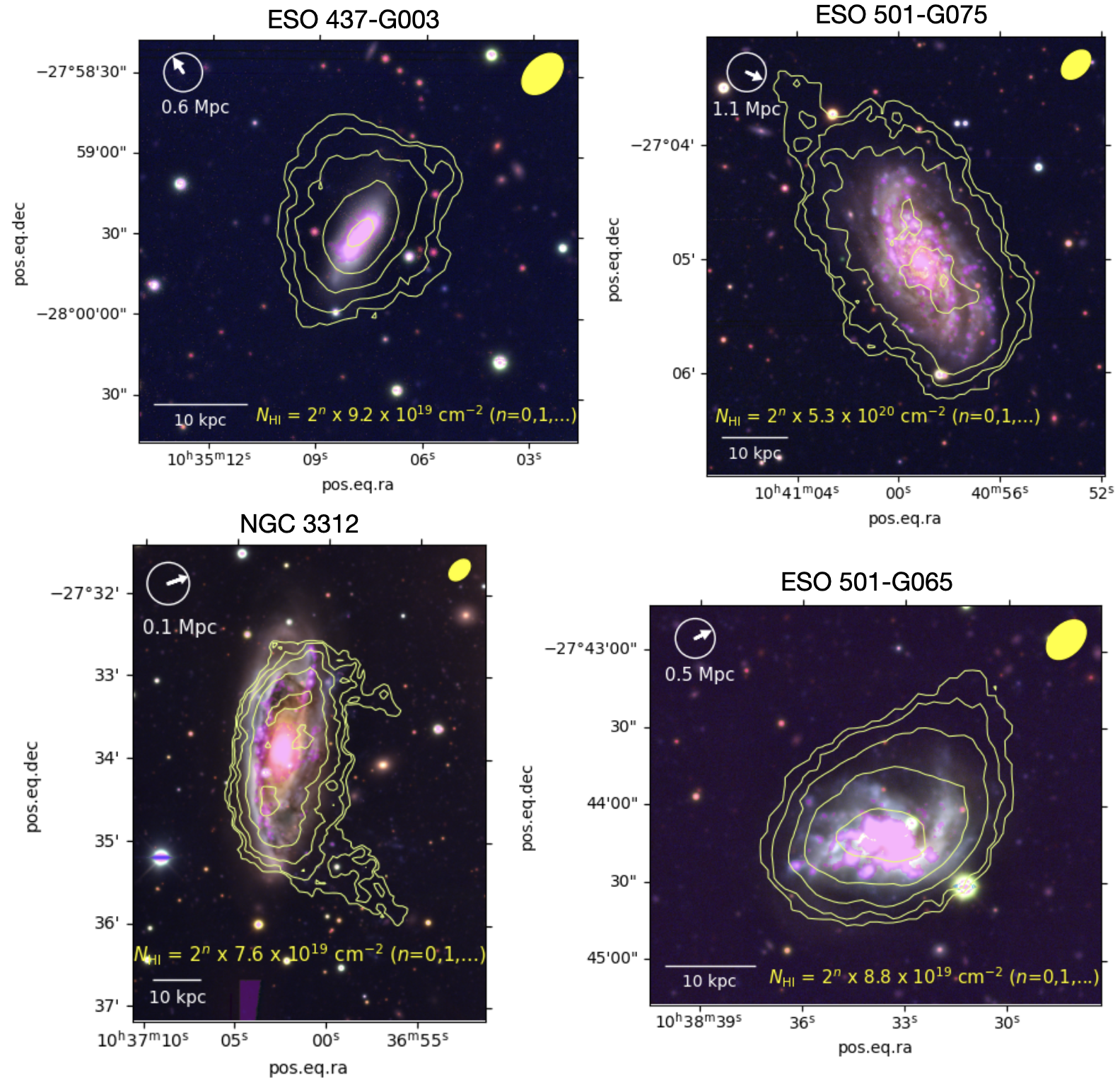}
    \caption{False rgb color images (red - r band, green - g band, blue - u band) overlaid with H$\alpha$ map (fuchsia) for 4 starburst galaxies in the H$\alpha$IR sample (see sec. \ref{sec:DMS}): ESO 437-G003, ESO 501-G075, NGC 3312, and ESO 501-G065. In yellow, MeerKAT HI contours for each galaxy, and the column density they represent. The beam size is displayed at the top right. The compass at the top left points in the direction towards the center of the cluster. The distance indicated below the compass is the projected distance of the galaxy to the center of the cluster. A scalebar of 10kpc can be found at the bottom left.\\
    }
    \label{fig:rgb_combo_intro}
\end{figure*}

\subsection{New imaging data}
\subsubsection{Optical broad and narrowband imaging: DECam}

We present in this work new, deep optical imaging of the Hydra I cluster, uniformly mosaicking the entire cluster in 5 broadbands (u, g, r, i, z,) and in the N662 narrow-band H$\alpha$ filter\footnote{\href{https://noirlab.edu/science/programs/ctio/filters/Dark-Energy-Camera}{noirlab.edu/}} from the Dark Energy Camera installed on the CTIO Blanco 4m telescope \citep[DECam;][]{flaugher2015dark}. Observations were carried out over 3 nights (2021-04-09 to 2021-04-11; project 2021A-0117; PI: R.~Kotulla).
With a surface brightness limit of $\mu( 3\sigma;10''\times10'')$=26.9 mag $\rm arcsec^2$ in the g band, they provide the optical emission necessary to estimate optical SFR and magnitudes.

Our optical measurements consist of 120 (119) u, g, r, H$\alpha$, (i, z) images, with a size of 23.3'x23.3', approximately 0.15 $\rm deg^2$. The observations were dithered to cover the whole cluster, and gaps in the CCDs with several different sets of observations. We mosaicked them into maps ranging between RA = [10:30:09.307, 10:50:33.567] and DEC = [-29.09.01.168, -25.48.22.051]. We studied the region within the limits RA = [10:31:41.0, 10:43:07.5] and DEC = [-26:00:44.0, -28:54:36.6] covering $\sim 7.5 \rm deg^2$ (see Fig. \ref{fig:map}). The S/N for the rest of the map was too low to produce any detection, since the outermost region was not covered with as many exposures.

The data processing comprised data reduction, normalization and scaling, and continuum subtraction. For the data reduction, we implemented the \texttt{obs\_subaru} package in the LSST science pipeline, including overscan and bias subtraction, flat-fielding, astrometric calibration relative to GAIA as reference, and photometric calibration relative to photometry obtained from PanSTARRS. 
For the background subtraction we used an adaptation of the original algorithm created by Hyper Suprime-Cam, instead of the standard detector-by-detector sky estimation. First, we normalized the data from multiple fields, and combined them to generate a template across the full focal plane. Then, we scaled in intensity to each individual frame, and performed the subtraction. The advantage of this approach compared to the standard one is that it preserves more extended structures that otherwise would be lost as background. The cost is slightly higher small-scale background residuals. Further details on the acquisition and processing of the data can be found in \citep{Hess2022}. To showcase the combined depth and quality of these measurements, Fig. \ref{fig:rgb_combo_intro} shows 4 galaxies from the H$\alpha$IR sample in the optical r,g,b and H$\alpha$ bands. Table \ref{tab:optical_data_intro_mags} contains the magnitudes extracted from them. We provide a summary of the techniques employed for the estimation of these magnitudes and their uncertainties in Appendix \ref{App1:optIm}.

\subsubsection{HI Imaging: MeerKAT}
HI spectral line observations were taken with the MeerKAT Radio Telescope \citep{Jonas16} during the first period of 4K Open Time (Project: SCI-20190418-CC-01, PI: C. Carignan). Data reduction for each 8 hour interval was performed employing the \texttt{CARACal} pipeline \citep{jozsa2020caracal,jozsa2020meerkathi} on the ilifu computer cluster hosted by the Inter-University Institute for Data Intensive Astronomy (IDIA). This pipeline applies the radio interferometry scripting framework STIMELA in order to handle open-source radio interferometry software packages (as CASA, Cubical, etc). 

The imaging of the HI spectral line involved creating a cube with WSClean. Only the horizontal (HH) and vertical (VV) polarizations of the sub-band 1370-1418 MHz (509-11030 km$s^{-1}$) covering the Hydra I cluster velocity range were reduced at the full frequency resolution (209 kHz, 44 km$s^{-1}$ at z = 0).
After smoothing and mosaicking, our final cube has a beam size of 11.8''$\times$18.0'' and rms noise of 0.13 mJy beam$^{-1}$ channel$^{-1}$, or a 1$\sigma$ HI column density sensitivity of N$_{\rm HI}$ = 3.0 $\times$ 10$^{19}$ cm$^{-2}$ channel$^{-1}$. 
Finally, we performed spectral line source finding using Source Finding Application \citep[SoFiA-2;][]{westmeier2021sofia}. We generated masks around 114 galaxies. Only 89 matched an optical detection in H$\alpha$ and are considered in this study. We refer to \cite{Hess2022} for a more in-depth description of the data processing and imaging.   HI flux was estimated within the SoFiA mask. 
 HI masses were estimated from flux luminosities following \cite{Meyer2017}.

\subsection{All-sky surveys}
\subsubsection{Mid-Infrared data: WISE} \label{S2:data_WISE}
In addition, we extracted infrared photometric measurements from the WISE Extended Source catalog from the brand-new calibration from \cite{cluver2024s}, and hence the latest
calibration that handles dwarf galaxies.  We derive stellar masses and star formation rates for the 196 galaxies in our sample from WISE. WISE's W1, W2, W3 and W4 bands were employed to estimate stellar masses and star formation rates for the galaxy members \citep{cluver2017calibrating}. These bands are centered on 3.4, 4.6, 12 and 22 $\mu m$ in the mid-infrared window, with a W1 point source sensitivity that reaches $\sim25$ $\mu Jy$ \citep[5 $\sigma$;][]{jarrett2016galaxy}.
As resolved extragalactic sources, these galaxies were characterized by WXSC native resolution stacked mosaics particularly constructed to compensate the poor angular resolution of WISE primary mirror and give account of either the galaxy or its surroundings.

 WISE is sensitive to emission from more evolved stellar populations and low temperature processes taking place in the galaxies' ISM and star-forming regions \citep{jarrett2012extending}, so it will offer a reference for our optical data, as well as a complementary chronicle of obscured star formation. 
 
 A possible source of contamination in our data is AGN activity. We employed the mid-infrared color-color relation to sample the AGN content of the galaxies, following Fig. 11 in \cite{jarrett2016galaxy}.  Galaxies with $\rm W2-W3$ color$>$0.8 are expected to contain an AGN or be ULIRGs. Only 2 galaxies (LEDA 751896, WISEA J103103.05-281826.1) show evidence for potential AGN contribution. Thus, we choose to disregard this factor for the sake of simplicity.

\subsubsection{Control sample: the AMIGA sample of isolated galaxies}
Projects such as AMIGA \citep[Analysis of the interstellar Medium of Isolated GAlaxies][]{verdes2005amiga} provide a valuable control sample of galaxies isolated from major interactions, required for any statistical study of the effects of environment on galaxy properties and evolution. The criterion of isolation of AMIGA implies that these galaxies are unlikely to have undergone a major interaction in the last 3 Gyr \citep{verdes2005amiga,verley2007amiga}, and reduced minor interactions (which have been quantified). Their interstellar medium and stellar component have been well characterized in their different phases across the optical \citep{verdes2005amiga}, far-infrared (FIR) luminosity derived from IRAS data \citep{lisenfeld2007amiga}, or radio-continuum \citep{leon2008amiga}. \cite{jones2018amiga} also provided HI scaling relations, including HI and mIR measurements. The AMIGA sample has been tuned over the years to supply with the most complete and largest sample of nearby isolated galaxies in the Northern hemisphere. Consequently, AMIGA galaxies have previously helped to demonstrate that galaxies in clusters evolve differently and more rapidly than galaxies in low-density environments \citep{espada2011amiga,martinez2012molecular,sabater2013frequency,scott2018abell,watts2020xgass}. We compare with their results to identify galaxies that are likely undergoing environmental quenching relative to those in isolation.

  \begin{table*}
     \centering
     \caption[]{Coordinates, redshift and optical absolute magnitudes of the galaxies in Fig. \ref{fig:rgb_combo_intro}.}
         \setlength{\tabcolsep}{3pt}
         \begin{tabular}{ccccccccc}
            \hline
            \noalign{\smallskip}
           
                &  RA & DEC & z &M$_u$ & M$_g$ & M$_r$ & M$_i$ & M$_z$ \\
            \noalign{\smallskip}
            \hline
            \noalign{\smallskip}
            ESO437$-$G003   & 10:35:07 & -27:59:32  &  0.0081 &  24.79$\pm$0.04  & 21.99$\pm$0.44  &  22.08$\pm$0.44 &  22.36$\pm$0.40  & 22.56$\pm$0.43 \\
            ESO501$-$G075     &  10:40:59 & -27:05:02  &  0.0127   & 22.89$\pm$0.42  & 19.80$\pm$0.42  &  19.59$\pm$0.48   &  19.68$\pm$0.51  & 19.88$\pm$0.51   \\
            NGC3312  & 10:37:01 & -27:33:54  & 0.0080  & 22.788$\pm$0.004  & 19.240$\pm$0.004  & 18.914$\pm$0.003  &   18.964$\pm$0.003  & 18.860$\pm$0.003   \\
            ESO501$-$G065     &  10:38:32  & -27:44:13  & 0.0077  & 23.08$\pm$0.39  & 20.25$\pm$0.35  & 20.31$\pm$0.37  &  20.58$\pm$0.43  & 20.78$\pm$0.43    \\

            \hline
         \end{tabular}

    {\textit{Column 1}: Name of the galaxies in Fig. \ref{fig:rgb_combo_intro}. \textit{Column 2}: Right ascension in hh:mm:ss. \textit{Column 3}: Declination in dd:m:ss. \textit{Column 4}: Redshift. \textit{Column 5}: u-band absolute magnitude. \textit{Column 6}: g-band absolute magnitude. \textit{Column 7}: r-band absolute magnitude. \textit{Column 8}: i-band absolute magnitude. \textit{Column 9}: z-band absolute magnitude. The full version of this table can be found online. \label{tab:optical_data_intro_mags}}
   \end{table*}

\begin{table*}
     \centering
     \caption[]{Star formation rates, stellar and gas masses of the galaxies in Fig. \ref{fig:rgb_combo_intro}.}
         \begin{tabular}{ccccc}
            \hline
            \noalign{\smallskip}
                &   $\rm SFR_{\rm H\alpha}[\rm M_{\odot}/yr]$ & $\rm SFR_{\rm IR}[\rm M_{\odot}/yr]$ & $\rm log(M_{\rm \star}[\rm \rm M_{ \odot}])$ & $\rm log(M_{\rm HI}[\rm \rm M_{ \odot}])$ \\
            \noalign{\smallskip}
            \hline
            \noalign{\smallskip}
            ESO437$-$G003     &  0.49$\pm$1.38 &  0.19$\pm$0.02  & 9.08$\pm$1.11  & 8.654$\pm$0.003 \\
            ESO501$-$G075  &  3.07$\pm$5.63 &  1.50$\pm$0.15  & 10.02$\pm$0.08  & 10.435$\pm$0.001\\
            NGC3312  &  12.45$\pm$2.40  & 2.69$\pm$0.27 &  11.12$\pm$0.08   &  9.318$\pm$0.001\\
            ESO501$-$G065     & 2.19$\pm$6.19  &  0.35$\pm$0.03 & 9.24$\pm$0.09  &  9.467$\pm$0.001 \\
            \hline
         \end{tabular}

    { \textit{Column 1}: Name of the galaxies in Fig. \ref{fig:rgb_combo_intro}. \textit{Column 2}: Star formation rates from H$\alpha$ narrowband. \textit{Column 3}: Star formation rates from IR. \textit{Column 4}: Logarithm of stellar mass from WISE. \textit{Column 5}: Logarithm of HI mass from MeerKAT measurements. The full version of this table can be found online.\label{tab:optical_data_intro_coords}}
    
 \end{table*}

 \subsection{Final sample}

Our final catalog has a total of 196 galaxies for which we are able to measure a SFR in either H${\alpha}$ or mIR. We will refer to this catalog as the H$\alpha$IR sample. For this sample, 89 redshifts come from MeerKAT, 90 redshifts were retrieved from optical spectroscopy of 6dFGS, and 17 from NED database (see Fig. \ref{fig:tabla1}).

For the optical selection, we applied a S/N>5 selection criteria. The lowest S/N in the sample is $\sim$6. We thus obtain SFRs for 186 galaxies, and lower limits for 10 galaxies (see Fig. \ref{fig:tabla1}). The origin of the lower limits is residual emission from nearby galaxies, low surface brightness, the presence of bright sources nearby, or an uncertain subtraction of the continuum (see App. \ref{App:lowlims}). In following sections we will trust the optical measurements to classify galaxies based on their star formation activity.

Out of the 196 galaxies in the sample, we measure SFRs from WISE in 150 galaxies, and upper limits in 46 galaxies. These upper limits are due to the lower sensitivity of the W3 band used to derive SFRs, or a relatively high stellar continuum emission in the region \citep{cluver2020galaxy}. 

The H$\alpha$IR sample covers the Hydra cluster out to a distance of 1.75$\rm r_{200}$, in the range of RA = [10:26:36, 10:46:16.8] in right ascension and DEC = [-30:18:00, -24:51:00] in declination. \cite{wang2020ram} showed that evidence of ram pressure stripping in HI disks of Hydra galaxies is dominant out to 1.5$\rm r_{200}$. Our coverage ensures to detect the earliest traces of environmental interaction with the gas content of our sampled galaxies.

\section{Results} \label{sec:SF}

   We present a catalog of optical broadband magnitudes, and H$\alpha$-based star formation rates for 196 galaxies in the Hydra Cluster out to 1.75 virial radii. 89 of these galaxies have HI masses from resolved MeerKAT observations (see Fig. \ref{fig:tabla1}). In the following subsections we describe the overall star-forming and gas properties of these 196 galaxies, that we refer to as the H$\alpha$IR sample.

\subsection{The star-forming main sequence}

  \begin{figure}
    \centering
    \includegraphics[width=\columnwidth]{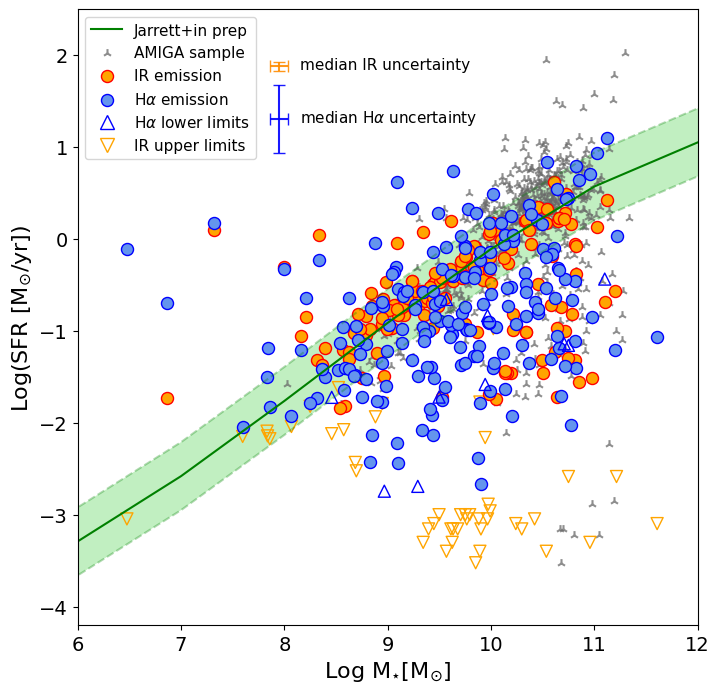}
    \caption{Star formation rate  versus stellar mass of the H$\alpha$IR sample in a logarithmic scale, estimated from the mIR (orange) and H$\alpha$ (blue) emission respectively. The optical and infrared SFR of galaxies with low confidence measurements are plotted as empty triangles. The green region is what we have defined as the Main Sequence (MS) of star-forming galaxies. We used Jarrett+ in prep fit of the MS and the $\sigma$ that \cite{bok2020h} employed to define their MS from WISE mIR observations of AMIGA galaxies. For reference, AMIGA isolated galaxies from \cite{bok2020h} are also shown in the background (gray). Average uncertainties for detections in H$\alpha$ and mIR have been provided at the top of the diagram for clarity.}
    \label{fig:MS}
\end{figure}

 The rates at which galaxies form stars are known to follow a tight correlation with stellar mass ($M_{\star}$) across environments, ages and cosmic times \citep[see e.g. ][]{cano2016spatially,davies2016gama,popesso2019main,popesso2023main}. This correlation is known as the Main Sequence of star formation for galaxies. The Main Sequence (MS) serves as reference to model the behavior of galaxies throughout their lifetime while they are mainly undergoing secular evolutionary processes. It therefore is an useful tool to identify deviations from this sequence. On one hand, infrared SFRs (SFR$_{\rm IR}$) offer a reference of past star-forming activity at timescales of about $10^{9}\rm yr$. On the other hand, SFRs estimated from the emission of the optical H$\alpha$ narrowband (SFR$_{\rm H\alpha}$) can provide insight on the SFRs on shorter timescales of $10^{7} \rm yr$. We employ SFR$_{\rm IR}$ to assess the quality of our optical measurements, while providing a notion of the star-forming activity of galaxies at longer timescales. We convert optical H$\alpha$ luminosities to SFR$_{\rm H\alpha}$ using Eq. (2) from \cite{kennicutt1998global}, and infrared luminosities to SFR$_{\rm IR}$ following Eqs. (4) and (6) from \cite{cluver2017calibrating}. Besides, stellar masses ($M_{\star}$) from WISE can not only offer insight on the history of gas processing of the galaxy, but also an internal dust correction for the SFR$_{\rm H\alpha}$. Our SFR$_{\rm H\alpha}$ were corrected for dust extinction following the stellar mass-dependent parameterization from Eq. (5) in \cite{garn2010dustcorrection} (see Appendix \ref{App1:optIm} for details). 

Figure \ref{fig:MS} displays the SFR vs $M{_{\star}}$ computed for the entire H$\alpha$IR sample. The blue points correspond to the SFR$_{\rm H\alpha}$ against the stellar mass, and the orange points correspond to the SFR$_{\rm IR}$. In both instances, the stellar mass is derived from WISE. Even though the scatter of the detected SFR$_{\rm H\alpha}$ is noticeably bigger than that of the infrared, both present similar overall behaviors across different mass ranges. 

In Fig. \ref{fig:MS} we define a region where galaxies are likely in the Main Sequence of star formation (green area). We employ the MS fit from Jarrett et al. in prep (green line in Fig. \ref{fig:MS}). This fit was derived using a large sample from the 2MRS, and thus accounts for dwarf galaxies. We build the Main Sequence region as the area within $1\sigma$ around the 2MRS's fit. This $\sigma= \pm0.37$ comes from the fit performed by \cite{bok2020h}, based on the comparison of pairs and AMIGA isolated galaxies using the WISE survey, and was employed to define their MS. \cite{bok2020h} fitted these galaxies employing a selection of those with stellar masses >$10^{8.5} \rm M_{\odot}$, to minimize the uncertainty of their study and mitigate the effect of missing dwarfs. However, their results are optimized for a mass range of $10^{9}-10^{11} \rm M_{ \odot}$. The mIR emission of isolated AMIGA galaxies measured by WISE is plotted in the background (gray) for reference. As we can observe in Fig. \ref{fig:MS}, the AMIGA sample is dominated by galaxies of stellar mass larger than $10^{10}\rm M_{\odot}$. Almost none live below $10^{8.5} \rm M_{\odot}$. The alignment of our sources with the MS is thus more robust at medium and slightly high stellar masses, which limits the interpretation of the Main Sequence at low stellar masses. 

In fact, the most under-represented galaxies in Fig. \ref{fig:MS} are low-mass galaxies that may have ceased or decreased their star-forming activity early on. From the 196 galaxies composing the H$\alpha$IR sample, 10 are lower limits in H$\alpha$ due to an elevated background signal or low flux from factors such as the presence of relatively bright sources along the line of sight to the source, or an oversubtraction of the continuum, as was detailed in Fig. \ref{fig:tabla1}. We display these lower limit values as blue triangles in Fig. \ref{fig:MS}. Similarly, 46 galaxies from the sample lack accurate measurements of their $\rm SFR_{IR}$. They can be identified as orange inverted triangles.

Low SFR populations with medium and low stellar masses are essential in any analysis of environmentally driven galaxy evolution, since it is widely accepted that environment increases the efficiency of a galaxy's quenching. However, there is a group of galaxies whose young and medium stellar population emission is above the MS. This suggests that in at least a subset of the medium-to-low-mass galaxies, star formation may be triggered by environment, rather than being quenched.

\subsection{Separating populations according to their SFR}\label{sec:DMS}


  \begin{figure*}
    \centering
    \includegraphics[width= 15cm]{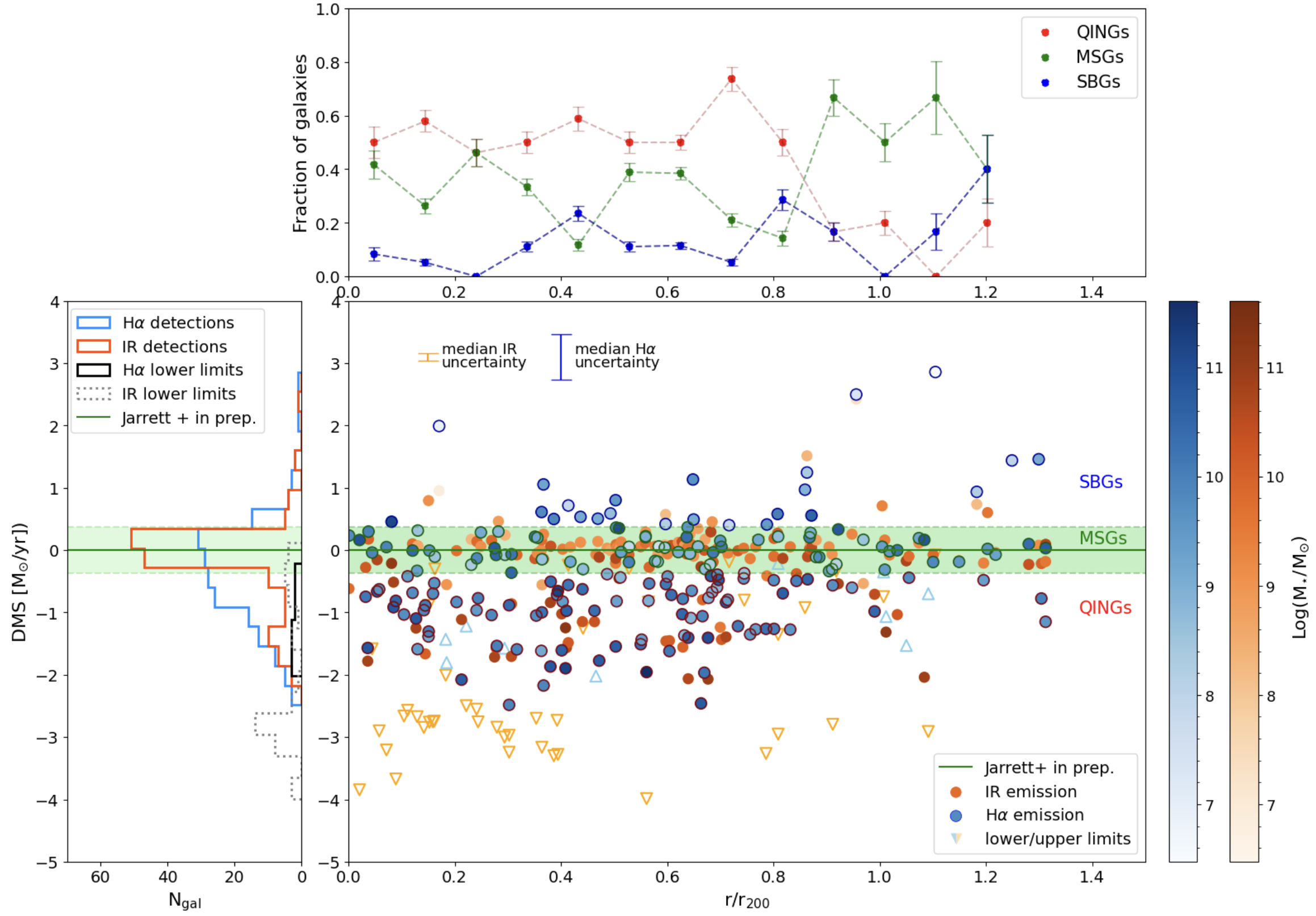}
    \caption{ Center: DMS as defined in Eq. \ref{eq:DMS}. That is, vertical distance between each SFR$_{\rm H\alpha}$ (blue) and SFR$_{\rm IR}$ (orange) marker, and the Main Sequence (MS), against the clustercentric projected distance. Empty triangles show the population of galaxies with lower (upper) limits in their SFR$_{\rm H\alpha}$ (SFR$_{\rm IR}$) estimations. All galaxies have been color-coded according to their stellar mass content. In addition, galaxies have been outlined in different colors in terms of their population: SBGs are marked in blue, MSGs in green, and QINGs in red.\\
    Left: Histogram of the DMS for the H$\alpha$ and mIR detections (blue and orange, respectively), and the H$\alpha$ and mIR lower/upper limits (black and gray, respectively).\\
     Top: Fraction of QINGs, MSGs, and SBGs (red, green and blue respectively) in distance bins of 0.96$\rm r/r_{200}$ out to 1.25$\rm r/r_{200}$.
    }
    \label{fig:DISTMShist}
\end{figure*}

We study in this section the offset of the galaxies from the MS as a function of the distance to the cluster center. We have chosen to employ the H$\alpha$ measurements rather than the infrared for this purpose because: (1) H$\alpha$ allows us to trace SFRs down to lower stellar masses than mIR (non-detections in WISE cannot actually be classified as quenched), and (2) SFR$_{\rm H\alpha}$ traces more recent star formation, and therefore it should be more closely linked to recent environmental processes affecting their morphologies and gas content.  These may be traced by the galaxies' morphology or gas content as seen in other studies \citep[e.g.][]{chung2009vla, poggianti2016jellyfish,serra2024meerkat}.

 We have computed the Distance to the Main Sequence (DMS) as the offset of the SFR of a galaxy to the $M_{\star}$ lying on the MS: 

\begin{equation}\label{eq:DMS}
 DMS_{\rm i} = SFR_{\rm i}-SFR_{\rm MS}.
\end{equation}

Figure \ref{fig:DISTMShist} displays the DMS in H$\alpha$ and IR, where the residuals provided by Eq. \ref{eq:DMS} are plotted against the normalized clustercentric projected distance (in units of $r_{\rm 200}$). SFR$_{\rm IR}$ are associated to the red color bar and the SFR$_{\rm H\alpha}$, to the blue color bar. Color accounts for stellar mass content. The errorbars are the uncertainties in SFR. The histogram to the left of Fig. \ref{fig:DISTMShist} shows the amount of galaxies at each DMS bin, separating detections and lower/upper limits. 
The only detections located below the MS are those with enough stellar mass to be inside the detection limit of WISE W3 band, or big enough to maintain certain levels of star formation after quenching. 
 
Figure \ref{fig:DISTMShist} classifies galaxies in different populations as a function of the distance from the MS. In this way, we can separate galaxies into populations defined by their star-forming activity, and understand where these different populations reside in the cluster. The three different H$\alpha$-based populations plus the lower limits from Fig \ref{fig:DISTMShist}, are:

\begin{enumerate}
    \item Quenching Galaxies (QINGs): 91 detections + 9 lower limits  ($\sim51\%$ of H$\alpha$IR galaxies) located below the MS ($SFR_{\rm H\alpha} < -1\sigma$). The 9 lower limits have uncertain corrections for the continuum, as discussed in App. \ref{App:lowlims}. They will not be taken into account in further discussions concerning SFRs.
    \item Main Sequence Galaxies (MSGs): 71 detections ($\sim37\%$ of H$\alpha$IR galaxies) on the MS (-1$\sigma<SFR_{\rm H\alpha}<1\sigma$). One galaxy has been flagged as an a lower limit. LEDA 770447 might have suffered an excessive subtraction of the continuum, and has been cataloged as lower limit. This galaxy will not be part of the following discussions.
    \item Starburst Galaxies (SBGs): 24 detections ($\sim12\%$ of H$\alpha$IR galaxies) above the MS ($SFR_{\rm H\alpha}>1\sigma$). 3 of these galaxies (ESO 436- G 036, ESO 501-G 017 and WISEA J103114.11-283955.4) have $SFR_{\rm H\alpha}<SFR_{\rm IR}$, which means they could be past their peaks of star formation. 
\end{enumerate} 

\section{Discussion} \label{sec:discuss}
 \begin{figure*}
    \centering
    \includegraphics[width=\textwidth]{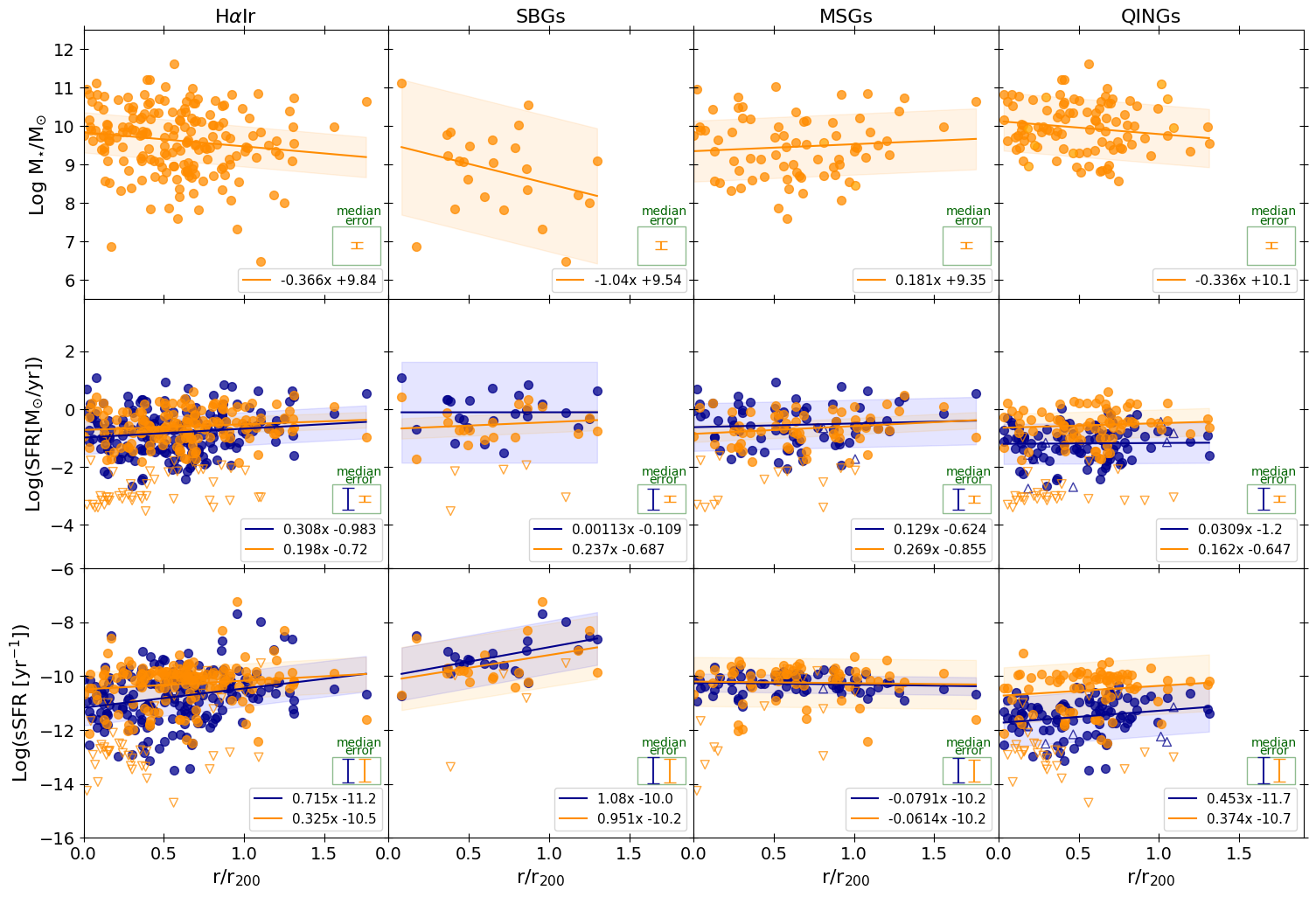}
    \caption{ From top to bottom and left to right: variation of $M_{\star}$, SFR and sSFR for the H$\alpha$IR sample, SBGs, MSGs and QINGs. Blue indicates H$\alpha$ measurements, and orange, mIR measurements. The legends contain the fits of the clouds, performed out to r/r200 = 1.25, since the sample is incomplete beyond that radius. The shaded areas are the 3$\sigma$ error of the regressions. The errorbars represent the median errors of the measurements (blue from $\rm H\alpha$ and orange from mIR) of $\rm M_{\star}$, $\rm SFR$, and $\rm sSFR$ for the populations considered in each column.}
    \label{fig:distvarios}
\end{figure*}

\subsection{The populations of the H$\alpha$IR sample}

Figures \ref{fig:MS} and \ref{fig:DISTMShist} show the star-forming activity of our sample at two different timescales: the more recent one ($\sim10^7$ yrs) portrayed by the optical SFR$_{\rm H\alpha}$, and a more historical account ($\sim10^{9}$ yrs) as measured by SFR$_{\rm IR}$. In the following sections, it is important to consider that we employ global star-forming tracers. Localized episodes of heightened or quenched star formation in a given galaxy are averaged over the whole galaxy. The changes in SFR$_{\rm H\alpha}$ that we report in timescales of tens of millions of years are global. 

At low $M_{\star}$, our H$\alpha$ and mIR based sample is unable to probe the full range of quenched galaxies. Faint, low-mass galaxies are upper/lower limits. These galaxies get quenched first and become more difficult to detect, so the depth of the WISE W3 band and even DECam's N662 filter is often not enough to differentiate the population that they belong to (SBGs, MSGs, QINGs). This implies that galaxies whose upper limits fall below the MS can be classified as quenched, but upper-limits in the SBG and MSG regions are ambiguous, yet limited to low stellar mass galaxies. The opposite occurs for lower limits.

In Figs. \ref{fig:MS} and \ref{fig:DISTMShist} we see that the high stellar mass population of galaxies in the cluster is dominated by quenching galaxies. They are the most abundant population of the H$\alpha$IR sample. We find 56 out of 91 detections, and 5 of 9 lower limits in the QING region to have $\rm SFR_{\rm IR}>SFR_{\rm H\alpha}$. This implies that recent star formation for these galaxies is lower than the average star formation that they experienced back in the timescale traced by IR. This is to be expected in quenching galaxies. Galaxies in the QING region with $\rm SFR_{\rm IR}<SFR_{\rm H\alpha}$ could be a result of the uncertainty in our estimations of SFRs (see Fig. \ref{fig:MS}), or alternatively very short and very recent episodes of slightly higher star formation. 

  SBGs are the group that are undergoing the most dramatic increase in star formation rate. Only ESO 436-G 036, ESO 501-G 017 and WISEA J103114.11-283955.4 could have passed their peak of star formation for having SFR$_{\rm H\alpha}$<SFR$_{\rm IR}$, as we explained in Sec \ref{sec:DMS}. In these three cases, we note that it is also possible that the galaxies suffer from a higher dust attenuation than the $\rm M_{\star}$-based model can account for. The other 21 SBGs have SFR$_{\rm H\alpha}$>SFR$_{\rm IR}$.  Of these, the SFR$_{\rm IR}$ for 3 falls in the SBG region, for 15 in the MS, and for 3 in the QING region. This implies that at least these 21 SBGs may have undergone a sharp increase in SFR only over the last $\sim10^7 \rm yrs$. 

\subsection{Star formation and projected distance to the cluster core} \label{subsec:sfprojdist}

  In this section we examine the link between the projected distance to the cluster core and the impact of environment in triggering bursts of star formation, or decimating the gas reservoir of galaxies until they get quenched beyond the detection limit. 
  
  In Fig. \ref{fig:DISTMShist} we showed how the different populations are distributed as a function of clustercentric distance. The top diagram displays how the fraction of SBGs, MSGs and QINGs changes with distance. The region beyond 1.25$\rm r_{200}$ is undersampled. It is not possible to draw conclusions from the distribution of the galaxies located in this area. We define 13 bins out to 1.25 $\rm r/r_{200}$. Each bin covers a region of 0.096 $\rm r/r_{200}$ instead of the more intuitive 0.1 $\rm r/r_{200}$ to ensure that the last bin does not cover an under-sampled area. We compute the fraction of each population by counting the number of galaxies within each bin and dividing the amount of SBGs, MSGs and QINGs in said bin by that number. The fraction of QINGS dominate out to $\sim$0.8$\rm r_{200}$, after which they decline.  The fraction of MSGs is the highest from 0.8$\rm r_{200}$ outwards, but contributes to the galaxy population all the way to the core. MSGs at the core are a group of massive, dusty ETGs including the BCGs, such as NGC 3311 \citep{richtler2020dust}. Finally, the SBGs' fraction increases up to $\sim$0.5 beyond $\rm r_{200}$. Within the virial radius, they maintain a low fraction, with a couple of increments at $\sim$0.4$\rm r_{200}$ and $\sim$0.8$\rm r_{200}$, respectively. 

 The previous observations suggest that there might exist a relation between the level of star formation and the distance to the cluster core, where environmental interactions are harsher and more frequent. However, we also need to consider the stellar mass of the galaxies as an account of their whole history of star formation. It is generally thought that lower stellar masses are associated with higher gas reservoirs, belonging to galaxies that have spent less time inside the coarse cluster environment. Still, we find low $M_{\star}$ galaxies close to the center, which strengthens the idea that their proximity to the cluster core is behind their increased star formation.
 
 In Fig. \ref{fig:distvarios} we investigate the relationship between the SFRs at the two different timescales represented by H$\alpha$ and mid-IR, and the projected distance for each population. We represent the stellar mass content, the star formation rate and the specific star formation rate ($\rm sSFR = SFR/M_{\star}$) vs. the projected distance to the cluster center normalized by the virial radius. We do so, from left to right, for the complete H$\alpha$IR sample, the SBGs, the MSGs and the QINGs. Blue color is associated with $\rm SFR_{H\alpha}$, and orange, $\rm SFR_{IR}$. Empty triangles are lower/upper limits. Detections were fitted employing the $\tt odr$ package from $\tt scipy$. This package allows to perform an orthogonal distance regression. Contrary to ordinary least squares, orthogonal distance regression takes into account the uncertainties of the variables to perform the regression. The choice to exclude lower limits from the fits introduces a bias towards more actively star forming galaxies, which is of little consequence to our analysis since we do not attempt to establish scaling laws or functions. In the following, we use this approach to compare infrared and optical star formation rates for SBGs, MSGs and QINGs, and see how they distribute over projected distance to the cluster core.  
 
 Figure \ref{fig:distvarios} shows how SBGs avoid the most central region of the cluster ($r/r_{\rm 200}<0.1$), where virialized galaxies dominate. According to the $M_{\star}$ vs $r/r_{\rm 200}$ diagram, SBGs are preferentially found within 0.7$r/r_{\rm 200}$. 
 In contrast, there is no gap between the location of the cluster core and the location of the closest QINGs, which span all throughout the cluster. This is consistent with observations showing that environmentally driven quenching extends from the cluster core out to $~2-3r_{\rm 200}$, well beyond the limiting radius considered in this study \citep[see][for an extensive review]{boselli2006environmental}. The results of \cite{lewis20022df} showed that galaxy transformation is driven mainly by processes effective even in lower density environments, even though galaxy evolution is highly impacted by other mechanisms such as harassment, tidal interactions or ram pressure stripping. Simulations also helped to explain how the quenching of central galaxies is more effective than for satellites. Gas removal can start at $2-3r_{\rm 200}$ \citep{zinger2018quenching} for satellite galaxies. Yet, the strongest gas loss is experienced at $r/r_{\rm 200}\leq0.5$ in these models. This is matched by the observed trend between increasing gas deficiency and decreasing distance from the cluster center \citep{solanes2001hi,boselli2006environmental,chung2009vla}.

The stellar content and the star formation rates of the general H$\alpha$IR sample do not seem to have a strong dependence on radial distance, although a slight trend can be observed, particularly for SBGs. This is at odds with other works that have found a connection between the quenching of star formation and distance to the cluster core \citep{raichoor2012galaxy,woo2013dependence,pintos2019evolution}. The presence of a number of gas-rich, star forming late type galaxies close in projection to Hydra's core \citep{Reynolds2021, wang2020ram, Hess2022} may be behind the weak trends manifested in Fig. \ref{fig:distvarios}.

However, when we compare the optical and the infrared fits of SFR and sSFR vs $\rm r/r_{\rm 200}$ for every population, we observe vertical offsets between H$\alpha$ and mIR that we interpret as indicative of timescales at which star formation happens in groups. In the case of SBGs, on average throughout the cluster, both SFR$_{\rm H\alpha}$>SFR$_{\rm IR}$ and $\rm sSFR_{H\alpha}$>$\rm sSFR_{IR}$ roughly verify, since they experience relatively recent increment in their star-forming activity. Regarding the MSGs, there is no evident  difference between H$\alpha$ and mIR derived vertical offsets as this family of galaxies is expected to have maintained their SFRs throughout the two sampled timescales. Finally, for the QINGs we find that SFR$_{\rm H\alpha}$<SFR$_{\rm IR}$ and $\rm sSFR_{ H\alpha}$<$\rm sSFR_{IR}$, which could suggest  significant quenching in the last tens of Myrs. However, many studies \citep[e.g.][]{zhang2023contribution,calzetti2024jwst} have shown how mIR tracers of SFR are contaminated by emission from older stellar populations. This contribution may have been particularly underestimated in the QING population. Stellar population synthesis would be needed to confirm whether the lower optical star formation is caused by a decrease in star-forming activity or the uncertainty in the estimation of mIR star formation. 

\subsection{Morphological and color classification}\label{subsec:morphcol}

\begin{figure*}
    \centering
    \includegraphics[width= 18cm]{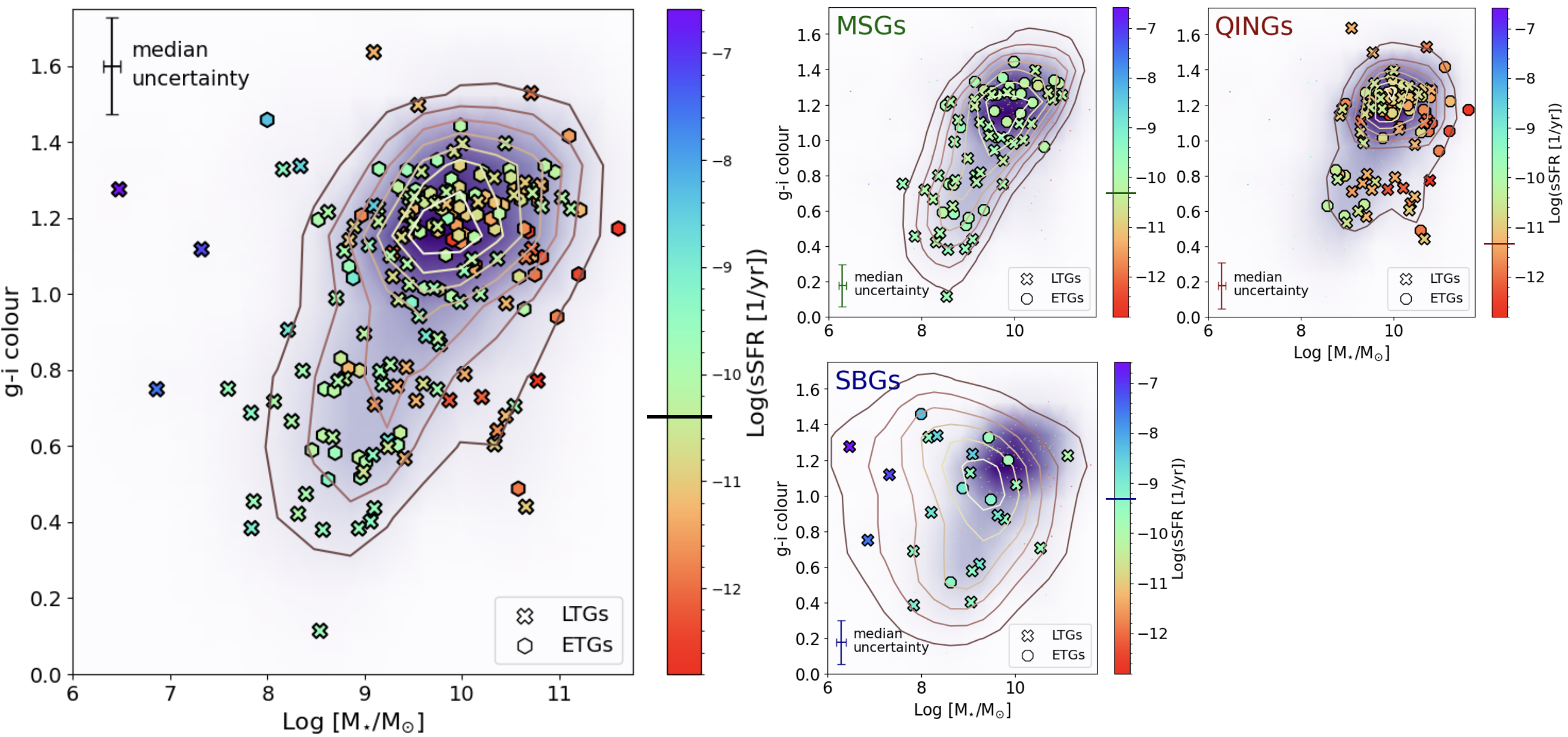}
    \caption{Color mass diagram for the complete sample of H$\alpha$IR galaxies. SBGs, MSGs and QINGs have been represented independently to the right of the diagram. LTGs are displayed as filled crosses and ETGs as hexagons.  The color code is representative of the sSFR$\rm _{H\alpha}$, and the horizontal line on the colorbar is the median sSFR of the sample. The purple shade follows the density of sources from the complete $\rm H\alpha Ir$ sample in every diagram. Contours indicate the density of the sample in each instance. The red sequence, the concentration of galaxies to the top right, is denser than the blue cloud to the bottom left, as would be expected in a galaxy cluster.
    \label{fig:CMD}}
\end{figure*}



   Before reaching a conclusion on the quenching state of the cluster, it is worth considering the selection bias of the sample that we are studying.

   Figure \ref{fig:CMD} displays the g-i CDM of the H$\alpha$IR galaxies, and the three different populations of SBGs, MSGs, and QINGs. The purple-scaled contours show the density of points of the complete sample, and the contours display the distribution of the group represented in each diagram. Galaxies have been color-coded by their sSFRs, and shaped depending on their morphology. We use the morphological information available in NED to separate galaxies into LTGs (spirals Sa-Scs and irregulars Irr), and ETGs (ellipticals E, lenticulars S0 and compact/peculiar c/Pec).

    The g and i bands are the least affected by dust emission, and allow us to see the color bimodality of the H$\alpha$IR galaxies. The red sequence (g-i$\geq$1) is more populated and has lower sSFRs than the blue cloud (g-i$\leq$0.8), but we still find a considerable number of relatively active LTGs and ETGs transitioning through the green valley ( 0.8$\leq$g-i$\leq$1). The large fraction of quenching galaxies in the sample is to be expected for galaxies in a cluster.

   We can see that the abundance of ETGs is higher for the quenching population, but some lie in the blue cloud as well. \cite{schawinski2014green} also finds a tail of ETGs in their u-r blue cloud for a redshift-and-magnitude-limited sample of SSDS galaxies. They suggested that blue ETGs recently transformed from LTGS into spheroids can transit more rapidly after star formation is quenched rapidly, explaining the presence of ETGs in the blue cloud. This would also help to understand the presence of LTGs with medium-low sSFR in the red sequence. Observations such as \cite{liu2019morphological}’s suggested that physical properties like SFR in cluster galaxies change faster than their structural properties, which was recently confirmed for the Hydra cluster by \cite{Lima-Dias2019}. A direct comparison between cluster and isolated ETGs and LTGs is out of the scope of this paper.

\subsection{Star formation and location inside the cluster} \label{subsec: asymmetry}
  To complement our study on the dependence on clustercentric radius, we discuss the location of galaxies in equatorial coordinates in Fig. \ref{fig:map}. The fuchsia and yellow squared regions in the map trace the extent of our optical and HI surveyed area respectively. HI detections are colored yellow. The H$\alpha$IR sample, represented by circular markers, has been separated in SBGs (blue), MSGs (green), and QINGs (red). QINGs populate the highest density region in the cluster core as shown as well in Fig. \ref{fig:distvarios}, but can also be found scattered across the rest of the cluster, particularly in the northern half. We find that 68$\%$ of QINGs are found at declinations greater than $\rm DEC > -27.75^{\circ}$, in the northern half of the cluster. 
  For the MSGs, 40$\%$ are north of the cluster core ($\rm RA_{core}=159.17^{\circ}$, $\rm DEC_{core}=-27.52^{\circ}$), and 60$\%$ to the south; while for the SBGs the fraction north-south is 33$\%$-67$\%$. 
 Thus we find a strong asymmetry in the distribution of galaxies in different stages of star formation in the Hydra Cluster. Starbursting galaxies preferentially reside in the south, while galaxies in the north are more quiescent. This north-south asymmetry in the distribution of star forming galaxies is suggestive of active feedback between the cluster and the cosmic web, in the direction aligned with the filament that connects Hydra to the southern Antlia cluster.

  \begin{figure}[h!]
    \centering
    \includegraphics[width=\columnwidth]{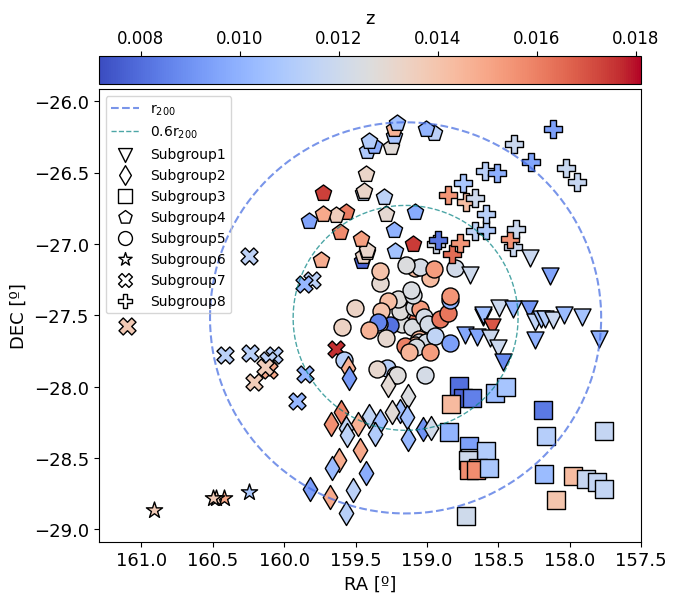}
    \caption{ Spatial distribution of the H$\alpha$IR sample, classified according to spatial distribution as well as redshift (subgroups 1-8) in section \ref{subsec:substructure}. The colorbar represents redshift: redder markers are more redshifted than bluer ones. The two dashed circles represent the virial radius, and the distance of 0.6$r_{200}$. We expect an overdensity of galaxies around the latter, given the higher number of galaxies that we can see in this region in Fig. \ref{fig:DISTMShist}.
    }
    \label{fig:maps_sub}
\end{figure}

\subsection{Substructure in the Hydra cluster} \label{subsec:substructure}
As we can observe in Fig. \ref{fig:map}, galaxies in Hydra I are apparently distributed in thread-like patterns that converge in the center of the cluster.  In Fig. \ref{fig:maps_sub} we show the classification of cluster galaxies separated into subgroups according to spatial distribution and velocity. We separate galaxies into subgroups 1 to 8, represented by different shaped markers. These subgroups trace the apparent threads that we observed in Fig. \ref{fig:map}. For the classification, we employed the Jenks Optimisation Method \citep{jenks1971error} to stablish the number of meaningful classes inside our sample as a function of redshift. Then, we used the {\tt scikit} python library to apply a K-means algorithm that clustered these groups in samples of equal variance in equatorial coordinates. The galaxies are color-coded by redshift. 

Previous studies have already observed that Hydra does not show high levels of substructure \citep{Lima-Dias2019,LaMarca2022}. \cite{Lima-Dias2019} used a sample of 193 Hydra galaxies from S-PLUS out to a virial radius, and found three possible substructures that would be located around coordinates $\rm (RA,DEC) =$ $(159.6^{\circ},-26.8^{\circ}),$ $(158.7^{\circ},-28.5^{\circ})$ and $(158.4^{\circ},-27.0^{\circ})$. 
\cite{LaMarca2022} analyzed the projected galaxy density distribution of 317 dwarf galaxies with $\rm M_r>-18$ out to 0.4 virial radius, which is the area where our classification is less robust, and found 2 overdensities. They propose that these overdensities are related to the two northern substructures of \cite{Lima-Dias2019}. The substructures that were found by these authors are aligned with subgroups 3, 4 and 8, respectively. Fig. 18 from \cite{Lima-Dias2019} presents a map similar to our Fig. \ref{fig:maps_sub}, where the size of galaxies is proportional to the probability that they belong to a substructure. The biggest galaxies trace the structures that we have labeled as subgroups 2 and 3, although they only find evidence for a substructure in the latter. Examining the supercluster environment around Hydra, threads 2, 3, 4 and 8 appear to be connected to well-defined filaments in the large scale structure.  A full analysis is beyond the scope of this paper, but the observed substructures are suggestive of galactic accretion along these filaments.

As we observed in Sec. \ref{subsec: asymmetry}, the asymmetry between star-forming and quenching galaxies in Hydra might be connected to an underlying larger scale structure feeding the cluster. New galaxies do not enter symmetrically, but preferentially from the south, and/or even from some of the small scale subgroups or threads observed in Fig. \ref{fig:maps_sub}. This view is further sustained by the fact that the Hydra cluster shares a filamentary connection with the southern Antlia cluster \citep{kraan1994optical,courtois2013cosmography}, that serves as a bridge between these two clusters, that are separated $\sim7.9^{\circ}$ in the sky. This filament feeds Hydra from the South \citep{Hess2015}, and could be associated to subgroups 2, 3, 6 and 7 in Fig. \ref{fig:maps_sub} \citep[see Figs. 10, 11, 12, 17, 18, 22 from][]{courtois2013cosmography}. This would help justify why the star formation activity is not intrinsically related to radial distance. It could also explain why less active galaxies are usually found in the northern half of the cluster, while those undergoing active or even bursting episodes of star formation are located to the south. Thus, the environmental transition from the filament to the cluster medium is a plausible origin for the SBGs increased star formation. In the following sections we examine the orbits in phase space and stablish a connection with their HI morphology, to strengthen the basis of this hypothesis.

\subsection{Infalling state} \label{subsec:ppsd}

 \begin{figure*}
    \centering
    \includegraphics[width=15cm]{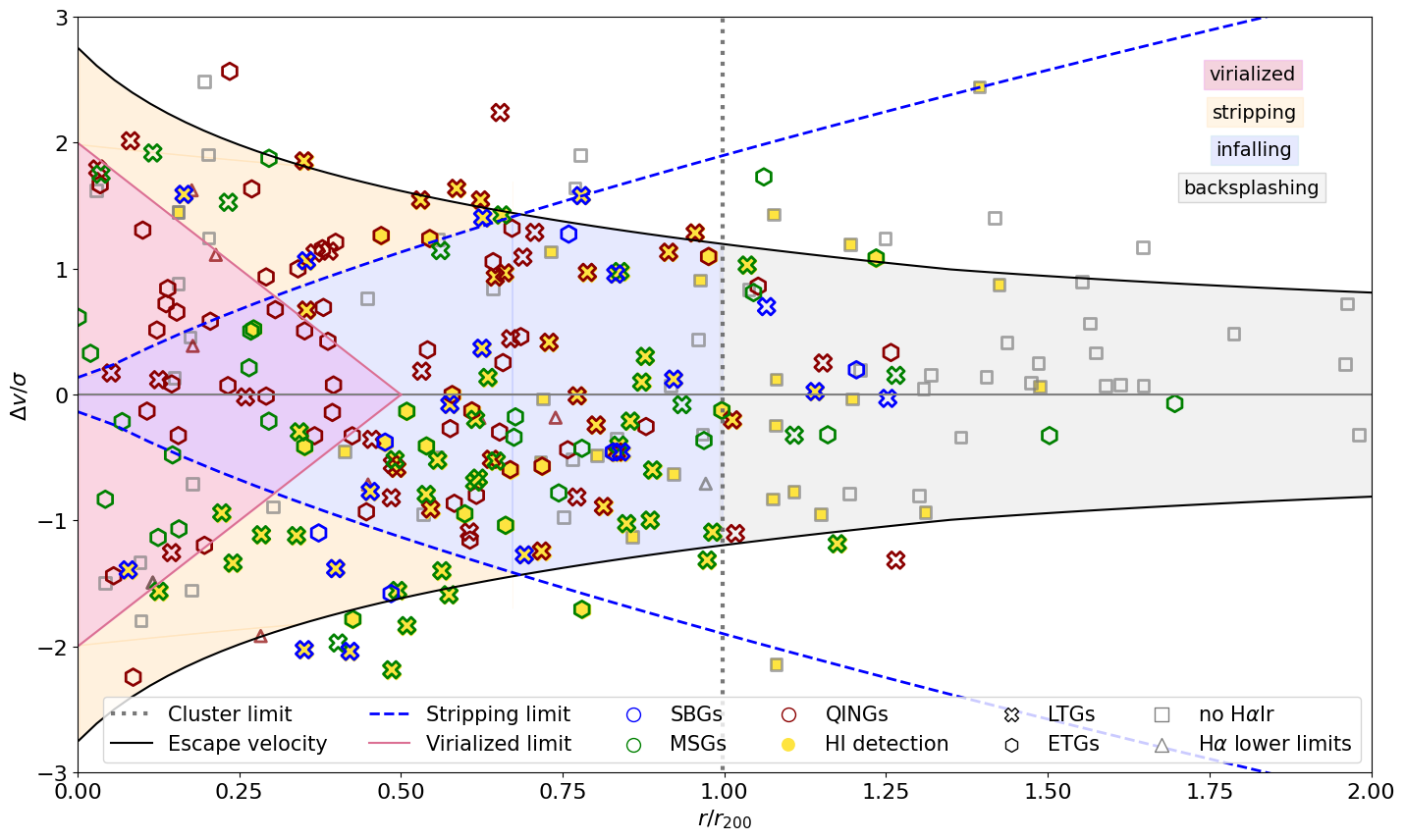}
    \caption{Projected Phase Space Diagram (PPSD) for H$\alpha$IR galaxies, showing the relation between velocity dispersion and clustercentric distance in the Hydra cluster. SBGs are represented with blue markers, while MSGs and QINGs are green and red, respectively. Hexagons represent the early type population, whereas crosses represent late type galaxies. Yellow-filled markers indicate HI detections. Black curves delimit the escape velocity region, whereas blue curves represent the stripping limit. Finally, vertical dashed lines show the virial radius. The area within these curves traces four distinct regions: the stripping region (yellow), where galaxies are being heavily stripped of their gas content; the virialized region (red) where virialized galaxies lie; the central infalling region (blue) where galaxies most distant from the velocity origin are probably crossing through their first orbits while the rest are already settled inside the cluster; and the backsplashing region (gray) where galaxies start falling into the cluster or are backsplashed from previous orbits.}
    \label{fig:PPSD}
\end{figure*}
  The orbital trajectories of galaxies can offer insight on the history of environmental mechanisms that may have impacted star formation in galaxies. However, tracing them is challenging since we lack information on the 3-dimensional location of galaxies relative to the cluster center. One possible approach is to infer the infalling state of galaxies in velocity space \citep{Rhee2017, smith2019PPSD,pasquali2019ppsd} using a Projected Phase Space Diagram (PPSD). Prior studies have shown that is possible to link the morphology of the cold gas reservoir to the location in phase space, and to the level of environmental impact on galaxies on other clusters \citep{jaffe2015budhies,yoon2017history}, as well as Hydra \citep{Wang2021,Reynolds2021}. 

The PPSD for our H$\alpha$IR sample is shown in Fig \ref{fig:PPSD}. The vertical axis is the recessional velocity of the galaxies normalized by the velocity dispersion of the cluster. The horizontal axis is the projected distance to the cluster center normalized by the virial radius. Positive values in velocity indicate faster receding galaxies (redshifted with respect to the center of the cluster), while negative values represent slower receding galaxies (blueshifted relative to the center of the cluster). The dashed vertical lines mark the virial radius. Galaxies to the left of the vertical blue line are expected to be gravitationally bound to the cluster. Black solid lines determine the escape velocity from the cluster at a given distance from its core, while blue solid lines show the stripping limit \citep{jaffe2015budhies}. For details of the models followed and parameters employed to trace these limits and the different regions of the PPSD for the Hydra I cluster, see Appendix \ref{App2:PPSD}.To visualize the galaxies that we mention in the rest of this discussion and their HI contours, see App. \ref{app:starburst}.
There are 4 different regions that have been emphasized in the diagram:


\paragraph{External infalling region (gray)}
Limited by $v_{\rm esc}$, the cluster and the infalling limits  \citep[$2.5r_{\rm 200}$;][]{oman2016satellite,Wang2021}, galaxies found here are in their first infall or ``backsplashing" \citep{Rhee2017}. Works like \cite{jaffe2015budhies,salinas2024constraining} have found that new cluster galaxies undergoing ram pressure stripping have tails of cold gas and optical material pointing away from the cluster core in first infall, and towards the core after first pericenter.
Among the four LTGs (1 QING, 2 MSG, 1 SBGs) and one ETG (a MSG) with HI detections in this area, only one shows an incipient HI tail. The SBG GALEXASC J103645.27-281004.4 (see Fig. \ref{fig:image4}) has a clear asymmetry in the direction opposite to the cluster core, characteristic of first infallers. The heightened SFR of SBGs as GALEXASC J103645.27-281004.4 could be due to the shock of plunging in the cluster. 

 \paragraph{Central infalling region (blue)}
Delimited by the cluster center, the escape and the stripping limit, in this area galaxies infalling for the first time converge with those that are backsplashing from previous orbits. Galaxies here can reach a stable orbit, after having passed pericenter up to a couple of times. Higher $\Delta v/\sigma$ values tend to be present in galaxies in their first orbits \citep{Rhee2017,yoon2017history}.

Most of the sample inhabits this zone, where we can find the most balanced relation between QINGs and MSGs+SBGs (51\%/49\% vs 28\%/72\% in backsplashing, 42\%/58\% in stripping, and 60\%/40\% in virialized regions). This makes it difficult to extract a precise quantifiable relation between environmental impact and distance that we can link to the bursts in star formation. The blueshifted half is more populated than the redshifted half, which conforms a pattern present in the whole diagram. 60\% of the sampled galaxies lie in the lower half of the diagram, while 40\% are in the upper redshifted half. As we proposed in sections \ref{subsec: asymmetry} and \ref{subsec:substructure}, H$\alpha$IR galaxies in Hydra seem to be infalling asymmetrically, preferentially from the southern half that is connected to the more blueshifted Antlia and even Centaurus clusters. Furthermore, in Fig. \ref{fig:maps_sub} we saw that most galaxies to the south (particularly in Subgroups 2 and 3) are blueshifted, and likely populating the central infalling region of the PPSD. This asymmetry in space and velocity distribution is also present for the HI detections. 

\paragraph{Stripping region (yellow)}
 Transition area between the cluster virial, escape and stripping limits. Inside this zone the anchoring gravitational force of the material is exceeded by the ram pressure. \cite{Wang2021} tried to predict the amount of galaxies undergoing ram pressure stripping (RPS) inside Hydra, and showed that most candidates for RPS from the WALLABY survey populate this region. This can manifest as harsh environmental erosion of their gas disks and reservoirs, disturbed gas disks, and long tails of lost material. We can find 24 HI detections in the stripping regions in our sample: 20 LTGs + 4 ETGs. According to this pattern, LTGs and ETGs with detected HI might be recent infallers at different stages of pre-processing and/or their first/second orbits.

\begin{figure*}
    \centering
    \includegraphics[width=15cm]{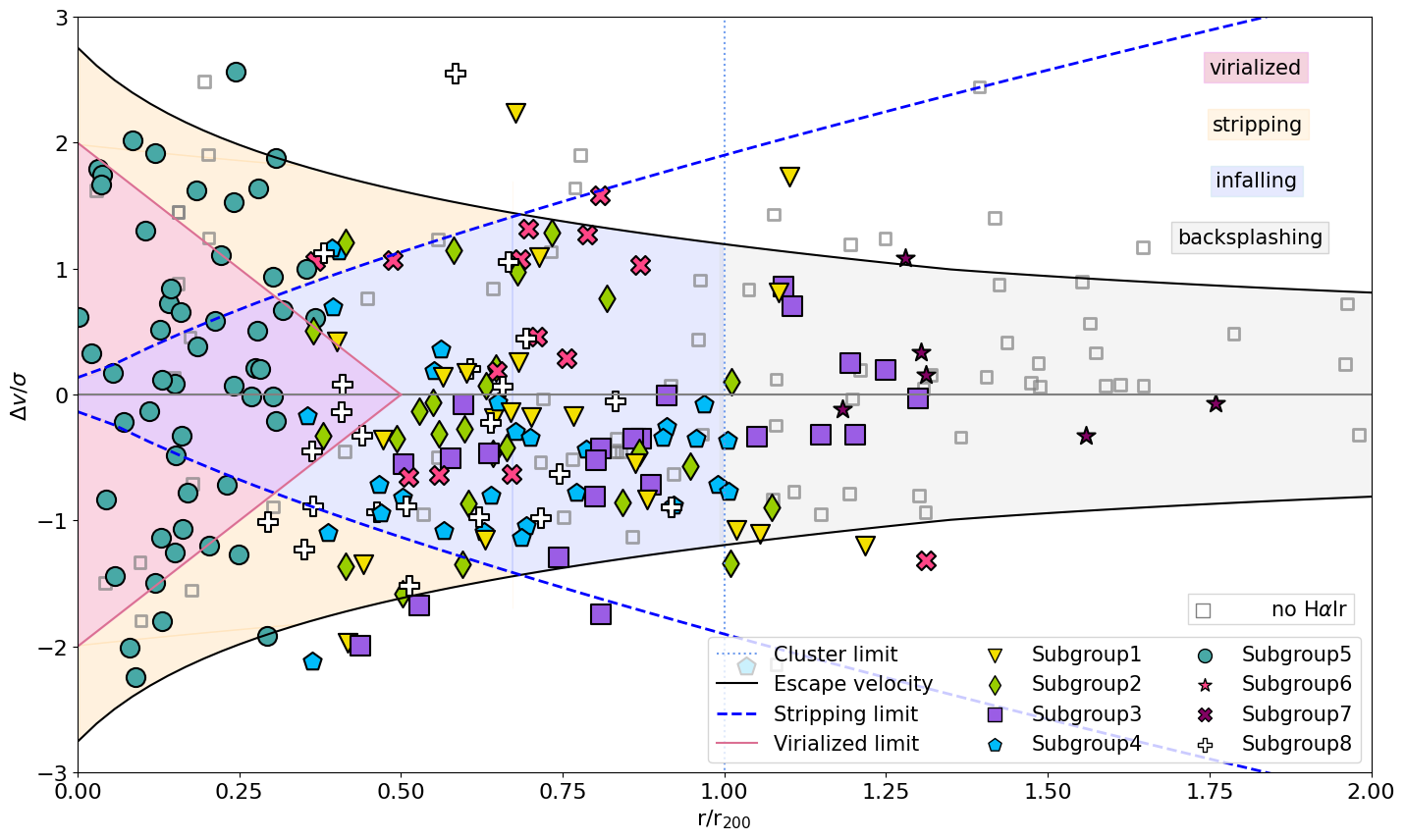}
    \caption{PPSD showing the distribution of galaxies that were associated to potential substructures in Fig. \ref{fig:maps_sub}. Different markers are associated to different subgroups. Empty markers are Hydra members outside of the $\rm H\alpha IR$ sample. Positive values in the y axis correspond to redshifted galaxies, and negative values to blueshifted galaxies, as explained in Fig. \ref{fig:PPSD}.} 
    \label{fig:PPSD_subs}
\end{figure*}

The redshifted half of the sample located in the stripping region is dominated by QINGs that retain some amount of star-forming activity even without a detectable atomic gas reservoir. Meanwhile, the blueshifted half of the galaxies in the stripping region is composed mainly of MSGs and SBGs, and contains most of the HI detections. It is possible that we are observing two different classes of galaxies: a redder one in g-i color, composed by galaxies that fell earlier in the cluster, and a bluer population of recently accreted galaxies. This trend was found as well in the Antlia cluster by \cite{Hess2015}.

 \paragraph{Virialized region (red)}
 The oldest cluster galaxies, primarily ETGs, reside here after virialization. The intracluster medium (ICM) plays a remarkable role in the suppression/enhancement of star formation in this area. Younger members in their first orbits will experience rougher hydrodynamical resistance from the ICM that sometimes manifests as long cold gas tails. The ram pressure stripped galaxy NGC 3312 (Fig. \ref{fig:rgb_combo_intro}) is a good example of how tail-like features and compressed leading edges in HI morphologies of SBGs could tell the story of backsplashed MSGs undergoing a recent period of increased star formation \citep{Hess2022}.

 \subsection{Substructures in phase space}

In Fig. \ref{fig:PPSD_subs} we display the location in phase space of the substructures that we classified in Sec. \ref{subsec:substructure}, to offer insight on the dynamics of accretion of galaxies from the cosmic web. 

\begin{itemize}
    \item Galaxies in Subgroup 1 are spatially distributed in a thread stretched from the core into the West direction. In velocity space, these galaxies assemble mostly in the blueshifted central infalling region. There is another group of redshifted galaxies with $\Delta v/\sigma>1$ that could be infalling for the first time.
    \item Subgroup 2 galaxies are infalling from the South of the Hydra cluster, likely from the filament connecting Hydra and Antlia. Most of them are in the central infalling region. Those with the higher absolute values of $\Delta v/\sigma$ and farther in projection to the core are probably infalling for the first time. Since many of these galaxies still maintain neutral gas reservoirs (Figs. \ref{fig:map} and \ref{fig:PPSD_subs}), they must not have been exposed to the cluster environment for long. We believe that they must be backsplashing in their first orbit or at the beginning of the second orbit at most. 
     \item Subgroup 3 contains mainly blueshifted galaxies in the southern half of the cluster. Their relatively high distance to the core gives the impression of being relatively new additions to the cluster. Despite the high dispersion in velocities between the galaxies in this group, their distribution in the PPSD suggest a common large-scale origin.
    
    \item Subgroup 4 galaxies are mostly in the blueshited central infalling region from the NW thread (according to Fig. \ref{fig:maps_sub}). We believe these galaxies could be connected to a filament that bridges Hydra and the Shapley supercluster.
    
    \item Subgroup 5 encompasses almost all galaxies in the virialized region, plus those galaxies closest to pericenter in the stripping region. This subgroup is quite heterogeneous: we can find elliptical galaxies such as NGC 3309 and NGC 3311 (Hydra's BCGs), and recently accreted spirals such as NGC 3312 that are past first pericenter \citep{Hess2022}.
    
    \item Subgroup 6 is formed by a group of redshifted, gas-rich galaxies dominated by the low surface giant ESO 437-44. In spite of their relative velocities with respect to Hydra's core, these galaxies seem to be infalling into Hydra for the first time through the filament that connects it to Antlia.
    
    \item  Members of Subgroup 7 extend to the East from the core. They populate the redshifted infalling (and partially stripping) regions with high $\Delta v/\sigma$. Outliers are likely chance alignments, that have been grouped due to their spatial proximity.
   
    \item Subgroup 8 behaves similarly to subgroup 3: most galaxies are in the blueshifted half of the central-infalling region, and some have even reached the virialized region. The rest are probably infalling for the first time and approaching or having already reached the stripping region for the first time, which will be reflected in their cold gas morphologies.  These galaxies connect the cluster to the NW with the cosmic web, in the direction where we can find the Cancer cluster.
    
\end{itemize}
  
This picture of the cluster shows an East-West asymmetry in addition to the north-south asymmetry observed in previous sections. Most recent infalls seem to come from the Eastern subgroups, in the direction where we found the connection to the Shapley supercluster, the Antlia cluster and the Centaurus cluster \citep[see][]{courtois2013cosmography}. Additionally, from Fig. 7 in \cite{courtois2013cosmography}, it is also evident that Hydra could share a filamentary connection with the Cancer cluster in the NW direction of Subgroup 8.

\subsection{Environmental impact upon gas} \label{subsec:himorph}

  \begin{figure*}[h!]
    \centering
    \includegraphics[width=18cm]{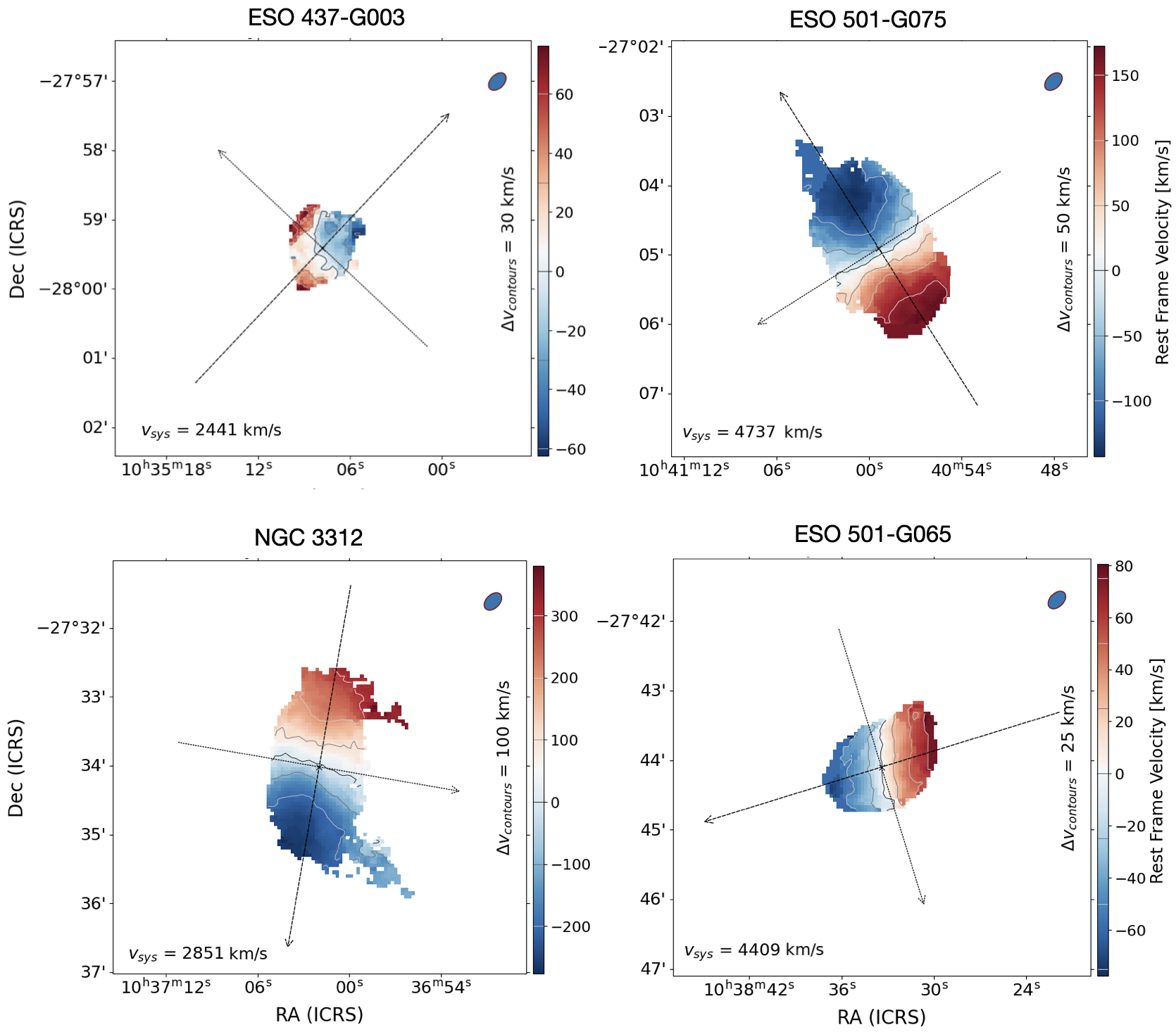}
    \caption{Moment 1 HI-velocity maps of the 4 starburst galaxies that were introduced in Fig. \ref{fig:rgb_combo_intro}, elaborated with \emph{SoFiA-Imaging-Pipeline (SIP)}. From left to right starting up: ESO 437-G003, ESO 501-G075, ESO 501-G009, ESO 501-G065. The rest of the starburst galaxies discussed in this work can be found in App. \ref{app:starburst}.\\
    }
    \label{fig:mom1_combo}
\end{figure*}

The final step to understand the link between environmental mechanisms and the heightened star formation activity of SBGs is to look for evidences of interactions in their gas morphology. Since gas is less gravitationally bound to galaxies than stars, it manifests hydrodynamical and gravitational interactions at earlier stages than the stellar component. In this section we examine the morphology of the neutral gas disks of the 18 SBGs with an HI detection, and compare them to the distribution of the star formation activity as traced by the H$\alpha$ emission. We focus our discussion on the effect of ram pressure stripping, which has previously been observed as far out as 1.25 $r_{200}$ in Hydra \citep{Wang2021}, although some models predict that it can begin as far out as $\sim$2–3 virial radii \citep{zinger2018quenching}.

Figure \ref{fig:rgb_combo_intro} displays the total intensity of the H$\alpha$ emission of four SBGs overlaid on their rgb images with the neutral gas contours delimiting the extent of the HI content down to column densities of $10^{19}\rm cm^{-2}$. In Fig. \ref{fig:mom1_combo} we present the moment 1 maps with the intensity-weighted velocity of the HI line. We use these 4 galaxies as an example. The full SBG sample with HI detection can be found in App. \ref{app:starburst}. The contour maps presented in Fig. \ref{fig:rgb_combo_intro} show the asymmetries in the HI disks of SBGs. 

We focus our discussion on the connection between HI asymmetries and ram pressure stripping, and how it could trigger bursts of star formation. However, we need to consider the possible impact of tidal interactions upon star formation rates. Tidal interactions could be particularly relevant in the ESO 436-44 and ESO 436-IG 042 system (\ref{fig:image1}), relative to ram pressure. Nonetheless, we still consider ram pressure to be playing a significant role in the star-forming enhancement  due to the neutral gas tails pointing north in the system. This suggests that ram pressure is likely acting on the group in addition to tidal effects. The angular separation of these galaxies in the sky is about 70’’, which we take as reference for an effective tidal interaction in Hydra. Among the rest of galaxies cataloged as SBGs, the one closest to a neighbor is ESO 501-G059, located 146’’ to the south of IC 2597. With more than double the distance between ESO 436-44 and ESO 436-IG 042, we consider that tidal interactions do not dominate over ram pressure when shaping the gas disks of the rest of SBGs.

It is challenging to link the strength of ram pressure acting upon a galaxy with their HI morphology using the orbital direction as a predictor. It is difficult to find the 3D movement of a galaxy, and there is a possible offset between the shape of the HI reservoir and rapid changes in the direction of movement. In addition, a quantification of the degree of asymmetry of an HI disk is highly dependent on technical factors such as signal to noise or resolution, as well as physical ones like inclination or surface brightness \citep{giese2016non,hank2025h}. A detailed account of all these factors is beyond the reach of this work. Instead, it is common practice to assume the most simple scenario where the HI tail (or the HI compressed contours) point in the opposite (same) direction of movement. Following this convention, in all SBGs regardless of their distance or direction to the cluster core the separation between HI contours gets smaller in the (presumably) leading edges. This suggests a steep increase in the gas density in the half that is leading the movement. The resulting higher pressure in these regions might be behind the recent burst of star formation in tailed galaxies with $\rm SFR_{H\alpha}>SFR_{IR}$. The trailing half tends to be elongated and have more distanced contours, which could be caused by the drag of the ICM \citep{chung2009vla,jaffe2018gasp,krabbe2024diagnostic}. Irregular features in the neutral gas disks have velocities compatible with the bulk rotational motions, indicating that the impact of environment is not strong enough to completely disrupt gas from the rotational pattern of its parent galaxy. Consequently, these galaxies probably have not lost enough gas yet to deplete their reservoirs in a way that decreases their star-forming activity. The heightened star formation of spiral SBGs is rather related to the gas compression in their leading edges, which is known to trigger and enhance star formation \citep{abramson2018case,bekki2013galactic}. For lower mass galaxies, the temporarily higher SFR is more likely caused by mass flows driven by moderate levels of ram pressure, as shown by \cite{zhu2023and}’s simulations, and observed for satellite galaxies in other clusters \citep{ebeling2014jellyfish}.

Many studies have previously considered the orbital direction of the galaxy as traced by HI morphology. They observed that radial orbits are usually associated with stronger RPS \citep{jaffe2015budhies,jaffe2018gasp,smith2022new,biviano2024radial, salinas2024constraining}. However, a galaxy does not need to be infalling radially to undergo RPS. From the SBGs in App. \ref{app:starburst}, only 4 galaxies show compressed contours and tails pointing away from the direction to the center of the cluster: LEDA 751896, ESO 436-G036, ESO 501-G017, NGC 3312. This is less than a quarter of the SBG sample. This fraction of galaxies with tails pointing away from the cluster core has also been observed by larger surveys in different wavelengths, such as \cite{salinas2024constraining}. 

The fact that  SBGs in general lie in the “recent-infaller” area of the PPSD \citep{Rhee2017} is compatible with previous results in the field \citep{yoon2017history,jaffe2018gasp,Reynolds2021}, showing that first infallers suffer star-forming enhancing shock upon first dive into the ICM. Starburst in galaxies triggered by the effect of environment have already been observed in other clusters such as Coma \citep{mahajan2010star}, Fornax \citep{serra2023meerkat}, or Virgo \citep{mun2021star}. In all these cases, environmental processes are easily identifiable by the asymmetries they produce in the gas disks of galaxies which are behind the sudden increase in star formation activity. More than half of our SBGs sample have detectable HI and manifest HI asymmetries easily identifiable by eye. 

 The asymmetries in spatial and phase space location that we showed in Figs. \ref{fig:map} and \ref{fig:PPSD} also manifest when studying the distribution of SBGs with detected HI. We find that 66$\%$ of SBGs with HI detections are located to the south of the cluster. Meanwhile, 75$\%$ SBGs are blueshifted, which rises to 83$\%$ for HI detections only. This further strengthens our theory that most of the SBGs might be recent infalls from a more blueshifted filament, potentially associated to subgroups 2, 3, 6 (Fig. \ref{fig:maps_sub}) to the south, or perhaps subgroups 4, 8 to the north. Therefore, the observed asymmetries in the disk morphology are likely connected to the recent sharp increase in their star formation, and heightened by pre-processing effects in the filament \citep{lee2021properties} and/or the shock of entering the cluster for the first time.

\section{Summary} \label{sec:summary}
The Hydra I cluster is an ideal system to study how galaxies transform in their coexistence with their neighbor galaxies and the ICM. In this work we use new DECam imaging and mIR WISE data combined with MeerKAT HI measurements to characterize the impact of environment on the star formation activity of the members of Hydra.
With a total of 196 redshift-selected galaxies with stellar masses ranging roughly between $10^{7.5}\rm M_{\odot}-10^{11}\rm M_{\odot}$ and spanning out to 1.75$r_{\rm 200}$, we have undertaken a multiwavelength study of their star-forming activity related to their stellar content, local environment, orbital state, color and morphology. The main results can be summarized as follows:
\begin{itemize}
    \item We have classified galaxies by their difference in SFR to galaxies in the Main Sequence with the same stellar mass. We find that $\sim51\%$ of galaxies are quenching (QINGs), $\sim37\%$ are still in the Main Sequence (MSGs), and $\sim12\%$ have undergone a starburst process in the last $10^7$ years (SBGs). 
    \item There is a N-S asymmetry in the distribution of star forming galaxies in the cluster. $32\%$ of QINGs, $60\%$ of MSGs, and $67\%$ of SBGs are located in the southern half of the cluster. Considering exclusively galaxies with detected HI gas reservoir, the percentages are $\sim65\%$ for QINGs, $\sim71\%$, for MSGs, and $\sim66\%$ for SBGs. More active and gas-rich galaxies are found preferentially in the southern half of the cluster.
    \item The complete sample has a balanced distribution of active and passive galaxies. We observe a fully grown red sequence, QINGs and MSGs with medium-low levels of sSFR. The green valley contains galaxies of the three different populations, supporting evolutionary scenarios of rapid and slow quenching. The presence of late types with lower levels of SFR in the red sequence and the green valley suggests that structural properties such as morphology change slower than physical properties \citep{liu2019morphological}. Finally, the blue cloud shows a population of active SBGs and MSGs, and some low sSFR QINGs. 
    \item An analysis of the phase space diagram shows that the distribution of Hydra galaxies is asymmetric in velocity. Most galaxies are blueshifted and not virialized, with the exception of some QINGs and a reduced number of MSGs and SBGs that are probably just transiting through the virialized region. Most HI detections are found to still be infalling, which hints to recent processes of assembly of the cluster. The location in phase space of potential substructure could suggest that the most recent galaxies fall preferentially from the Eastern half of the cluster. 
    \item Another remarkable feature is that the 18 SBGs with HI detections have disturbed gas disks. We have connected the asymmetries to the bursts in star formation and hypothesized that most of them might be galaxies that joined the cluster recently. In particular, they could be entering from the southern Hydra-Antlia filament and have these starbursts powered by pre-processing or the shock of the first dive into Hydra.
\end{itemize}

While past studies show little substructure in Hydra, our detailed look at the star forming and gas properties of galaxies reveal substantial asymmetry, which likely reflects the recent accretion history of the cluster. In view of our results, we deduce that Hydra has a dual nature. On one hand, it shows a population of mature and aged galaxies that follow the Main Sequence or that have already moved away from it. These galaxies are located preferentially to the northern half of the cluster, and suggest some level of substructure while remaining mostly out of the virialized region in phase space.
On the other hand, we find a population of recently accreted galaxies undergoing a recent boost in their star formation. These galaxies might be in their first orbits in the cluster after being newly accreted from the cosmic web or a filament, particularly the Antlia-Hydra filament. These two complementary natures make Hydra a unique structure at large scale, both evolved and in active process of accretion of new material, and a promising laboratory for future multiwavelength studies.

\paragraph{Data availability}
The data displayed in Tables \ref{tab:optical_data_intro_mags} and \ref{tab:optical_data_intro_coords} are available in electronic form at the CDS via anonymous ftp.

\begin{acknowledgements}

The authors would like to acknowledge the tragic passing of co-author, Dr Thomas Jarrett, whose support and guidance were essential at every stage of this work. We would like to express our sincere gratitude to the anonymous referee for their valuable feedback, that has significantly strengthened the methodology and results of this work. Author C.C.C. would like to thank Amidou Sorgho, Manuel Parra, Emanuela Pompei and Yara Jaffé for useful discussion. Authors C.C.C., K.M.H, L.V.M.,M.-L.G.-M. and R.I. acknowledge financial support from the grant CEX2021-001131-S funded by MICIU/AEI/ 10.13039/501100011033, and from the grant PID2021-123930OB-C21 funded by MICIU/AEI. Author C.C.C additionally aknowledges financial support from the grant TED2021-130231B-I00 funded by MICIU/AEI and by the EuropeanUnion NextGenerationEU/PRTR. R.C.K.K. gratefully acknowledges partial funding support from the National Aeronautics and Space Administration under project 80NSSC18K1498, and from the National Science Foundation under grants No 1852136 and 2150222. H.C. was supported by National Key R\&D Program of China NO. 2023YFE0110500, and by the Leading Innovation and Entrepreneurship Team of Zhejiang Province of China (Grant No. 2023R01008). T.H.J. acknowledges support from the National Research Foundation (South Africa). M.E.C. acknowledges the support of an Australian Research Council Future Fellowship (Project No. FT170100273) funded by the Australian Government. S.B.D. acknowledges financial support from the grant AST22.4.4, funded by Consejer\'ia de Universidad, Investigaci\'on e Innovaci\'on and Gobierno de España and Uni\'on Europea –- NextGenerationEU, also funded by PID2020-113689GB-I00, financed by MCIN/AEI.
J.S.G., is thankful for funding of this research provided by the University of Wisconsin-Madison College of Letters and Science. RCKK thanks the South African National Research Foundation for their support. Author C.C.C. acknowledges the Spanish Prototype of an SRC (SPSRC) service and support funded by the Ministerio de Ciencia, Innovación y Universidades (MICIU), by the Junta de Andalucía, by the European Regional Development Funds (ERDF) and by the European Union NextGenerationEU/PRTR. The SPSRC acknowledges financial support from the Agencia Estatal de Investigación (AEI) through the "Center of Excellence Severo Ochoa" award to the Instituto de Astrofísica de Andalucía (IAA-CSIC) (SEV-2017-0709) and from the grant CEX2021-001131-S funded by MICIU/AEI/ 10.13039/501100011033 \citep{spsrc_cita}.
The MeerKAT telescope is operated by the South African Radio Astronomy Observatory, which is a facility of the National Research Foundation, an agency of the Department of Science and Innovation.
Part of the data published here have been reduced using the CARACal pipeline, partially supported by ERC Starting grant number 679627 “FORNAX”, MAECI Grant Number ZA18GR02, DST-NRF Grant Number 113121 as part of the ISARP Joint Research Scheme, and BMBF project 05A17PC2 for D-MeerKAT. Information about CARACal can be obtained online under the URL: \href{https://caracal.readthedocs.io}{https://caracal.readthedocs.io} \citep{jozsa2020caracal}.
We acknowledge the use of the ilifu cloud computing facility – www.ilifu.ac.za, a partnership between the University of Cape Town, the University of the Western Cape, Stellenbosch University, Sol Plaatje University and the Cape Peninsula University of Technology. The ilifu facility is supported by contributions from the Inter-University Institute for Data Intensive Astronomy (IDIA – a partnership between the University of Cape Town, the University of Pretoria and the University of the Western Cape), the Computational Biology division at UCT and the Data Intensive Research Initiative of South Africa (DIRISA).
This work made use of the iDaVIE-v (immersive Data Visualisation Interactive Explorer for volumetric rendering) software (DOI – 10.5281/zenodo.4614116 – \href{https://idavie.readthedocs.io}{https://idavie.readthedocs.io}.
This work made use of the CARTA (Cube Analysis and Rendering Tool for Astronomy) software (DOI 10.5281/zenodo.3377984 –  https://cartavis.github.io).
This research has made use of the NASA/IPAC Extragalactic Database (NED), which is operated by the Jet Propulsion Laboratory, California Institute of Technology, under con- tract with the National Aeronautics and Space Administration.This research made use of \emph{Photutils}, an Astropy package for detection and photometry of astronomical sources \citep{larrybradley2024}
      
\end{acknowledgements}

\bibliographystyle{aa}
\bibliography{aa53831-25_arxiv.bbl}

\begin{appendix} 

\section{Photometry and magnitudes} \label{App1:optIm}

  \begin{figure*}[h!]
    \centering
    \includegraphics[width=18cm]{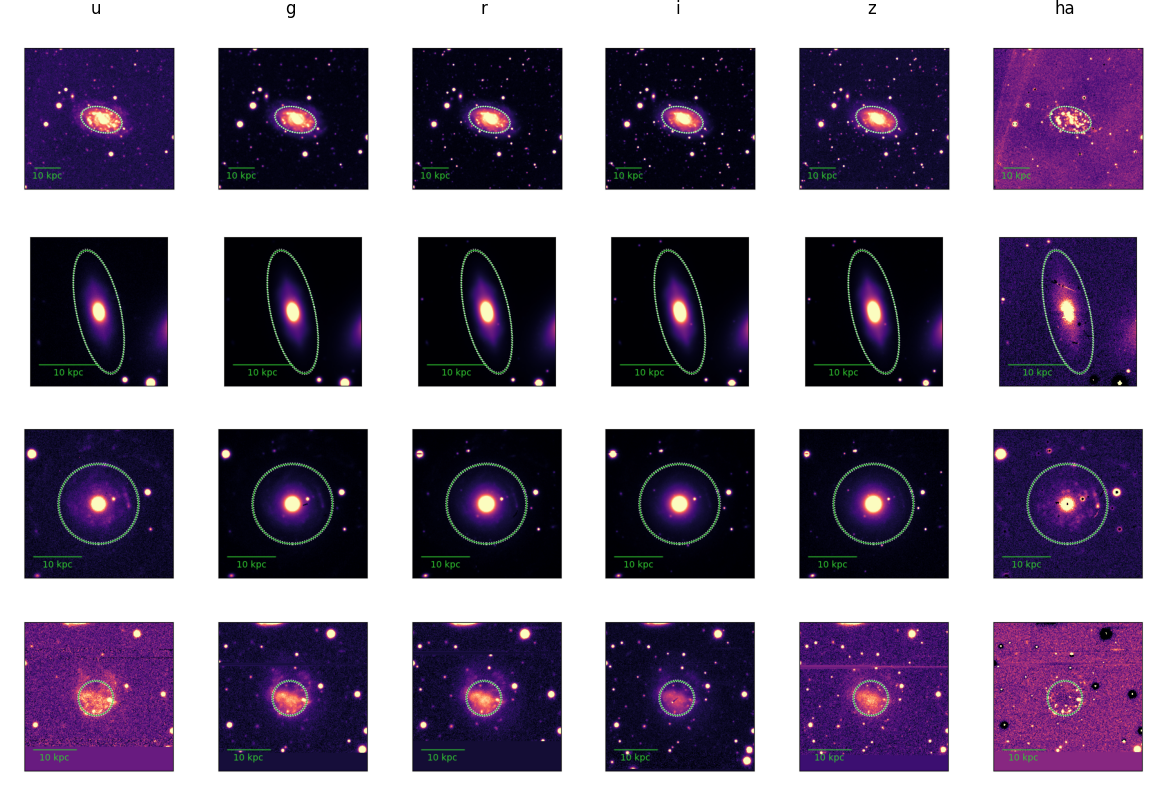}
    \caption{r26 isophotes defined for the photometry extractions in each band for four galaxies with different morphologies. From left to right: u, g, r, i, z broadbands and the continuum-subtracted H$\alpha$ narrowband. From top to bottom: ESO 436-G038, ESO 437-G013, ESO 437-G044, GALEXASC J103804.55-284333.9. The horizontal green line at the bottom left of every image represents a scale of 10kpc. The image cutouts were chosen, when possible, to avoid the presence of other galaxies in the images, while balancing the need to include enough background to produce meaningful statistics. Contaminating bright sources were sigma-clipped and not taken into account in the estimation.\\
    }
    \label{fig:photometry}
\end{figure*}
In this section we explain how the optical data was treated, including the continuum subtraction of the H$\alpha$ maps, and how the photometry and magnitudes were extracted for the sample. 

\subsection{Continuum subtraction}
Continuum subtraction of the H$\alpha$ narrow-band mosaic involved the scaling of N662 and r images to the same instrumental zeropoint. The resulting line-free image $I_{\rm H_{\alpha CS}}$ is obtained by subtracting the broad-band ($I_{\rm r}$) from the narrow-band image as follows ($I_{\rm H_{\rm \alpha}}$):
   
   \begin{equation} \label{ContinuumSubtraction}
   I_{\rm H_{\alpha}CS} = I_{\rm H_{\alpha}} - \frac{I_{\rm r} \times f_{\rm w_{\rm r}} - I_{\rm H_{\alpha}} \times f_{\rm w_{\rm H_{\alpha}}}}{f_{\rm w_r} - f_{\rm w_{H_{\alpha}}}} ,
   \end{equation}
   
\noindent where $f_{\rm w_{\rm r}} = 1276 \r{A}$ and $f_{\rm w_{\rm H_{\alpha}}} = 170 \r{A}$ are the filter widths for both continuum red emission and H$_{\rm \alpha}$ emission. The information about DECam's filters is publicly available in NOIRLab's website\footnote{\href{https://noirlab.edu/science/filters?field_telescope_target_id=538&field_instrument_target_id=547}{noirlab.edu/science/filters/}}.

 We note that, from equation \ref{ContinuumSubtraction}, one source of uncertainties that must be taken into account is the difference in filter width end centering of each band. For very blue sources we slightly underestimate H$\alpha$, in the range of g-r=0 - 0.4 we are within +/- a few percent, and towards redder sources we slightly overestimate the real H$\alpha$-flux. The most extreme case, for a 14 Gyr old elliptical model reaching g-r of 0.7, this overestimate is around 7\% of the uncorrected H$\alpha$-flux. From g-r = 0.4 - 0.7 the error is essentially linear, so at g-r$\sim$1 the uncertainty should be around 10\%. We choose to disregard this contribution, which amounts to less than 5\% of the total error budget discussed in more detailed in App. \ref{App1:uncertainties}.

\subsection{Source finding}
The location of galaxies with ionized gas emission in the maps was performed through a rigorous and strict visual inspection of the narrow-band, continuum subtracted blind map of the Hydra I cluster. The detection threshold was at S/N>5 , being the lowest S/N in the sample of S/N = 6.11. 

\subsection{Photometry}
We employed the \textit{Python} package {\tt photutils} \footnote{\href{https://photutils.readthedocs.io/}{photutils.readthedocs.io}} to define elliptical regions from which we measured the signal. We removed bright sources not belonging to the galaxy with the {\tt sigma\_clip} utility. Each measurement's uncertainty is a combination of both the photometric and the background uncertainties, the latter dominating the final uncertainty estimation. They are estimated respectively as the error computed from the {\tt aperture\_photometry} utility, and as the standard deviation of the background signal that was subtracted to correct the image. 

We implemented automated r26 isophote detection in the r band to perform the photometry, following \cite{chamba2022edges} and references therein. In Fig. \ref{fig:photometry} we
show a random selection of galaxies that illustrates the range of galaxy properties (clumpiness, concentration, crowded fields, etc), and the areas where we measured their light. This demonstrates the light distribution for galaxies of different morphologies and how the uncertainties that we estimated (see Fig. \ref{fig:CMD}) account for these variations. ESO 436-G038 (RA, DEC = 10h33m53.8253s, -27d49m42.261s) is a LTG with active star formation in its arms, that shows throughout the different bands. ESO 437-G013 (RA, DEC = 10h36m53.9904s, -27d55m01.092s) is a ETG with no substantial variation in the light distribution across different bands. ESO 437-G044 (RA, DEC = 10h41m42.3187s, -28d46m47.237s) is an LTG galaxy with faint arms that do not contribute substantially to the total measured light, so we have selected only the brightest region at the core. Finally, GALEXASC J103804.55-284333.9 (RA, DEC = 10h38m04.6008s -28d43m38.28s) is an irregular galaxy that shows a strong variation in its light profile across different bands.

Once the signal $\rm S$ was measured, the conversion into flux was performed in the following way: 

\begin{equation}\label{eq:flux_bands}
    f_{\rm band} = S\times10^{-\frac{zpt}{2.5}}\times M_{\rm AB}^{zero}\times \frac{FWHM}{CWL} ,
   \end{equation}

\noindent where $zpt$ is the zeropoint for a given band, $M_{\rm AB}^{zero}$ is its AB magnitude at the zeropoint, FWHM (Full Width at Half Maximum) is the spectral width of the band, and CWL (Central Wavelength) is the wavelength of the line. Once the fluxes have been estimated, colors for two given bands i and j can be calculated from their magnitudes:

\begin{equation}\label{eq:mag_ur}
    M_{\rm ij} = 2.5(log\phi_j -log\phi_i) -(E{j}-E_{\rm i})-K_{\rm corr}
   \end{equation}

\noindent being $E_{\rm ij}$ the internal extinction of the filters. The K-correction is taken from \cite{schawinski2014green} as $K_{\rm corr}=0.05$. All the  parameters involved in the process are shown in Table \ref{tab:photparam}. 

\subsection{Dust correction for star formation}

Star formation was converted from H$\alpha$ luminosities following Eq. (2) from \cite{kennicutt1998global}, under the assumption that "the ionizing radiation comes from young stars that formed at a constant rate over $\approx10 \rm Myr$ in the disc of ionized gas.

To correct SFR from dust extinction we applied the parameterization implemented by \cite{garn2010dustcorrection}. They applied polynomial fits to 120650 star-forming galaxies from SDSS DR7 (according to a BPT diagram), and S/N > 3 in all emission lines but O[III] (S/N>2). This resulted in Eq. \ref{eq:dust_cor}:

\begin{equation}\label{eq:dust_cor}
    A_{\rm H\alpha} = \sum_{\rm i=0}^{n} B_i X^i
   \end{equation}

\noindent where $A_{\rm H\alpha}$ is the extinction in the H$\alpha$ band, $X = log_{\rm 10}(M_{\star}/10^{10}\rm M_{\odot})$ accounts for the stellar mass content, and  $B_0=0.91$, $B_{\rm 1}=0.77$, $B_{\rm 2}=0.11$, $B_{\rm 3}=-0.09$ are the parameters of the fit. This correction is only based on the stellar mass of the galaxy, which for our sample is derived from WISE.

As reported by \cite{garn2010dustcorrection}, equation \ref{eq:dust_cor} predicts the extinction of a galaxy with a given stellar mass to within a typical error of 0.28 mag, which is comparable to the accuracy with which extinctions can be estimated from the Balmer decrement for the model sample. They report that residuals tend to be higher at higher stellar masses, as a consequence of downsizing once the most massive galaxies quench, and the relative weight of extinction in relation with the reduced Ha emission grows. However, the authors showed that the model is robust regardless of the selection bias of the sample, and can offer an statistical estimation of the extinction from the lowest stellar masses to the highest without substantial variation in the distribution of the residuals.

 \begin{center}
            \begin{table*}[h!]
            \centering
            \caption{\label{table:flux_par}Parameters employed for the estimation of fluxes (eqs \ref{eq:flux_bands}, \ref{eq:mag_ur}).}
            \begin{tabular}{ p{2cm} p{1cm} p{1cm} p{1cm} p{1cm} p{1cm} p{1cm} p{1cm}} 
            \hline
               \hspace{0.25cm}Parameter & \hspace{0.3cm} u  & \hspace{0.3cm} g & \hspace{0.3cm} r & \hspace{0.3cm} i & \hspace{0.3cm} z & \hspace{0.25cm} ha & \hspace{0.2cm}Ref.\\
             \hline
            
               \hspace{0.5cm}zpt [\AA] & \hspace{0.1cm}$28.45$ & \hspace{0.3cm}$27$ & \hspace{0.3cm}$27$ & \hspace{0.3cm}$27$& \hspace{0.3cm}$27$& \hspace{0.3cm}$27$ & \hspace{0.2cm} [1] \\
            
               \hspace{0.2cm}$M_{\rm AB}^{zero}$ [Jy] & \hspace{0.15cm}$1545$ & $1003.70$ &  $1276.20$ &  $1281.00$ & 
               $1289.30$ & \hspace{0.1cm} $145.4$ & \hspace{0.2cm} [2] \\
             
               \hspace{0.3cm}$CWL$ [\AA] & \hspace{0.1cm}$3650$ & $4770.80 $ & $6371.30 $ & $7774.2 $ & $9157.90 $ & \hspace{0.25cm}$6563 $ &\hspace{0.2cm} [1] \\
             
                \hspace{0.2cm}$FWHM$ [\AA] & \hspace{0.2cm}$880$ & $1003.70$ & $1276.20$ & $1281.00$ & $1289.30$ & \hspace{0.1cm}$145.4$ & \hspace{0.2cm} [1] \\
            
                \hspace{0.75cm}$E$ & \hspace{0.15cm}$0.62$ & \hspace{0.15cm}$0.36$ & \hspace{0.15cm}$0.36$ &\hspace{0.15cm}$0.45$ & \hspace{0.15cm}$0.45$ & \hspace{0.15cm}$0.45$ & \hspace{0.2cm} [3]\\
             \hline
            \end{tabular}\\
            {From left to right: parameter, value in u band, value in g band, value in r band, value in i band, value in z band, and reference from where they where extracted. References: [1] \href{https://noirlab.edu/science/filters?field_telescope_target_id=538&field_instrument_target_id=547}{noirlab.edu}, [2] \cite{abazajian2009seventh}, [3] \cite{kim2007reddening}.}\label{tab:photparam}
            \end{table*}
      \end{center}  

\subsubsection{Uncertainties derived from photometry} \label{App1:uncertainties}
The errors related to the estimation of the signal through photometry comprise:
    \begin{itemize}
        \item The uncertainty of the sum of the signal within the r26 aperture, propagated by the total error of the image. This error is estimated by combining the standard deviation of the image with the Poisson noise of sources. The Poisson error of the sources is derived from the gain of the DECam camera, which can be consulted in \href{https://noirlab.edu}{noirlab.edu}. For more details, we suggest consulting the documentation of {\tt $\rm calc\_total\_error$} and {\tt ApertureStats} classes of {\tt photutils}.
        \item The second one is the statistical error of the process, computed as the standard deviation of the background signal that we subtract to correct the image. This is the dominant source of uncertainty for most galaxies. It incorporates local background variations, and accounts for the uncertainty introduced by the continuum subtraction.
        \item The uncertainty introduced in the determination of the r26 isophote. This is estimated as the dispersion between the signals for the isophote at 26 mag/arcsec2 in the r band, and the 2 closest isophotes before and after this level. The weight of this uncertainty is higher for smaller galaxies.
    \end{itemize}

The uncertainties related to the conversion from signal to SFRs were performed as the square root of the sum of the squares of the errors in the quantities involved in  Eq. (2) from \cite{kennicutt1998global} and \ref{eq:dust_cor}. The same approach was taken for the estimation of magnitudes with Eq. \ref{eq:mag_ur}.

\subsection{Selection of lower limits in the sample} \label{App:lowlims}

\begin{figure}[h!]
    \centering
    \includegraphics[width=\columnwidth]{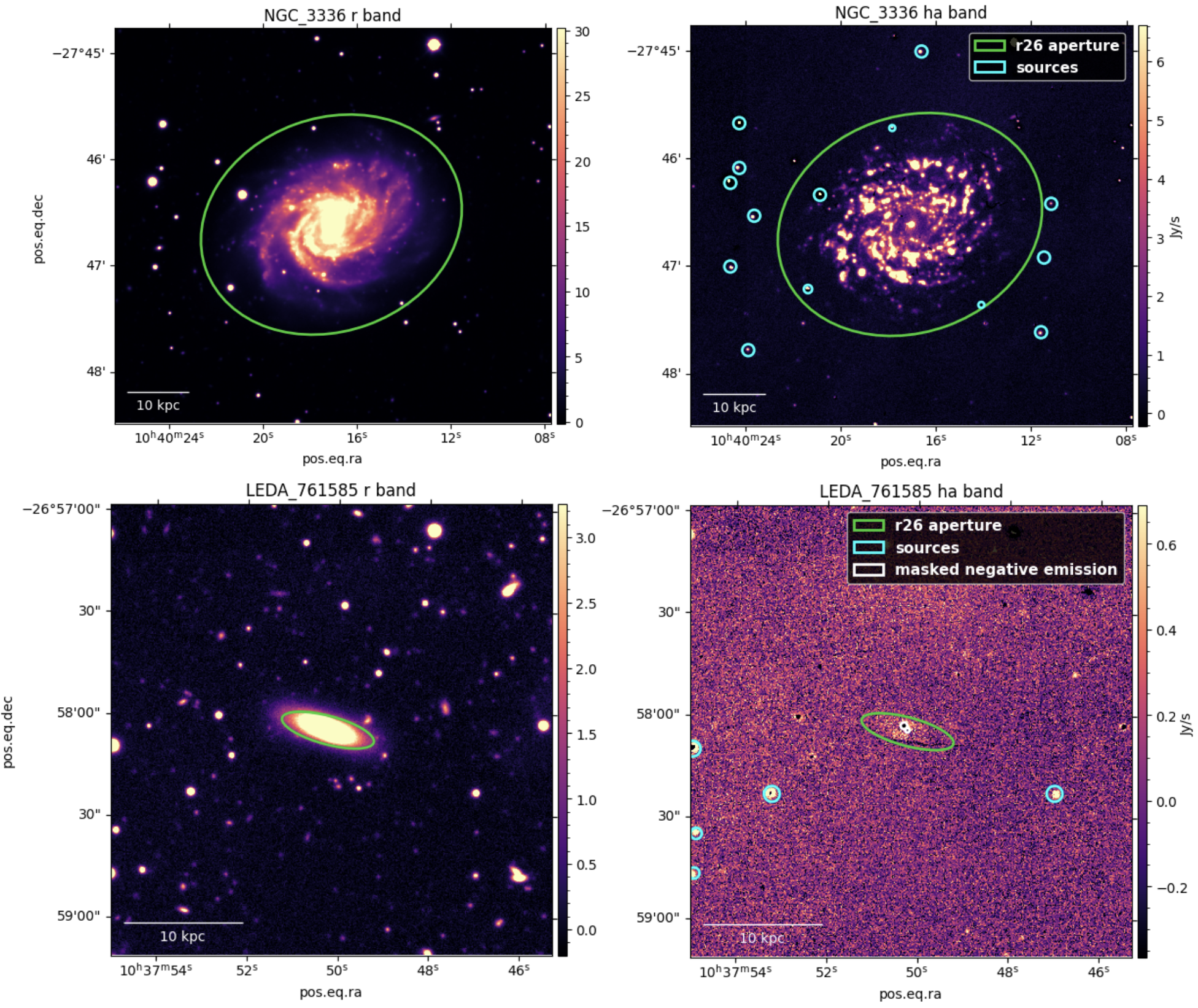}
    \caption{r and H$\alpha$ emission of a detection with homogeneous and relatively low background noise (NGC 3336, top), and a lower limit with uncertain continuum subtraction and inhomogeneus background noise distribution (LEDA 761585, bottom). In particluar, LEDA 761585 is the galaxy with the lowest S/N in the sample (S/N $\simeq$ 6). The green ellipse represents the r26 magnitude isophote in which we performed the photometry. The white regions had negative emission due to uncertain coontinuum subtraction, and were masked and set to background emission levels. Idem for the blue regions, masking bright stars.\\
    }
    \label{fig:lowlims_app}
\end{figure}

Figure \ref{fig:lowlims_app} shows an example of a detection and a lower limit. The lower limits strictly have S/N > 5, but the continuum subtraction is uncertain and the background noise is relatively high and inhomogeneus inside the aperture.

%
\section{Projected Phase Space Diagram} \label{App2:PPSD}
In this section we describe the steps we followed to build the velocity limits for the phase space diagram of the Hydra cluster. 

\subsection{Escape velocity}
The escape velocity from the cluster at a given distance from its core was estimated employing a virial mass of $M_{\rm 200} = 3.02\times 10^{14} \rm M_{\odot}$ distributed within a virial radius $\rm r_{200} = 1.35 Mpc$ according to a NFW potential  \citep{Navarro1996,jaffe2015budhies} following Eq. \ref{eq:PPSD0}. The paramters K, s and $\rm g_{c}$ are given by Eqs. \ref{eq:PPSD1}, \ref{eq:PPSD2}, \ref{eq:PPSD3}. The speed of light is c, G is the gravitational constant, and $r_t=6.1$ kpc is the truncation radius as inferred by \cite{jaffe2015budhies}.

\begin{equation} \label{eq:PPSD0}
 v_{\rm esc}=\left\lbrace\begin{array}{c} \sqrt{\frac{2GM_{\rm 200}K}{3r_{\rm 200}}}\hspace{1cm}r < r_{\rm 200} \\ \sqrt{\frac{2GM_{\rm 200}}{3r_{\rm 200}s}}\hspace{1cm}otherwise \end{array}\right. 
\end{equation}

\begin{equation} \label{eq:PPSD1}
 K = g_c\left(\frac{ln\left(1+cs\right)}{s}-ln\left(1+c\right)\right) +1 
\end{equation} 

\begin{equation} \label{eq:PPSD2}
 s \simeq \frac{\pi}{2} \frac{r_t}{R_{\rm 200}} 
\end{equation}

\begin{equation} \label{eq:PPSD3}
 g_c = \left[ ln\left(1+c\right) - \frac{c}{\left(1+c\right)}\right]^{-1} 
\end{equation}

\subsection{Stripping limit}
The stripping limit, that is, the limit at which the anchoring gravitational pull $\Pi_{\rm gal} = 2\pi G \Sigma_s \Sigma_g$ is exceeded by ram pressure $P_{\rm ram} = \rho_{\rm ICM}v_{\rm gal}^2 $, can be computed following \cite{jaffe2015budhies}:

\begin{equation}
v_{\rm gal}= \sqrt{\frac{2\pi G \Sigma_s \Sigma_g}{\rho_{\rm ICM}}}
\end{equation}

where $\Sigma_s$ and $\Sigma_g$ are the stellar and gas disc density profiles, respectively. The surface densities are given by $\Sigma=\Sigma_0 e^{-\frac{r}{R_d}}$,
$\Sigma_0=\frac{M_d}{2\pi R_d^2}$. The ICM gas density profile follows a $\beta$-model: $\rho_{\rm ICM}=\rho_0[1+(\frac{r\pi}{2R_c})^2]^{-3\beta/2}$.

Tables \ref{table:RP3} and \ref{table:RP4} contain the parameters employed in the calculus of $P_{\rm ram}$, and $\Pi_{\rm gal}$ \citep{jaffe2015budhies}.

\begin{center}
            \begin{table}[ht!]
            \centering
            \caption{\label{table:RP3}Parameters used for the calculus of $\rho_{\rm ICM}$.}
            \begin{tabular}{ |p{3cm}|p{3cm}|  } 
            \hline
               \hspace{0.75cm}Parameter & \hspace{1cm}Value \\
             \hline
               \hspace{1.25cm}$\beta$ & \hspace{1.25cm}$0.5$ \\
        
               \hspace{1.24cm}$\rho_0$ & \hspace{0.45cm}$14.7\times10^{-3}$ $cm^{-3}$ \\
           
               \hspace{1.2cm}$R_c$ & \hspace{1.1cm}$65$ kpc \\
             \hline
            \end{tabular}\\
            {$\beta$ is the $\beta$-model parameter of the ICM gas density profile, $\rho_0$ is the central initial density of the ICM, and $\rm R_{c}$ is the core radius fitted by \cite{rizza1998x} for the cluster A9631. They serve as a model to estimate the stripping limit in phase space.}
            \end{table}
      \end{center}

      \begin{center}
            \begin{table}[ht!]
            \centering
            \caption{\label{table:RP4}Parameters used for the calculus of $\Pi_{\rm gal}$.}
            \begin{tabular}{ |p{3cm}|p{3cm}|  } 
            \hline
              \hspace{0.75cm}Parameter & \hspace{1cm}Value \\
             \hline
              \hspace{0.9cm}$M_{\rm d,stellar}$ & \hspace{0.6cm}$4.6\times10^{10} \rm \rm M_{ \odot}$ \\
           
              \hspace{0.9cm}$M_{\rm d,gas}$ & \hspace{0.1cm}$\simeq 0.06\times M_{\rm d,stellar}$ \\
            
              \hspace{0.9cm}$R_{\rm d,stars}$ & \hspace{0.9cm}$2.15$ kpc \\
         
              \hspace{0.9cm}$R_{\rm d,gas}$ & \hspace{0.5cm}$ 1.7\times R_{\rm d,stars}$ \\
             \hline
            \end{tabular}\\
            {$M_{\rm d,stellar}$ and $M_{\rm d,gas} $ are the stellar and gaseous mass of the disc where we model the galaxy’s anchoring pressure $\Pi_{\rm gal}$. $R_{\rm d,stars}$ and $R_{\rm d,gas}$ are the stellar and gaseous radius for this disc.}
            \end{table}
      \end{center}

 \section{Starburst galaxies}\label{app:starburst}

 
 \begin{figure*}[h!]
 \begin{subfigure}{\columnwidth}
 \includegraphics[width=17cm]{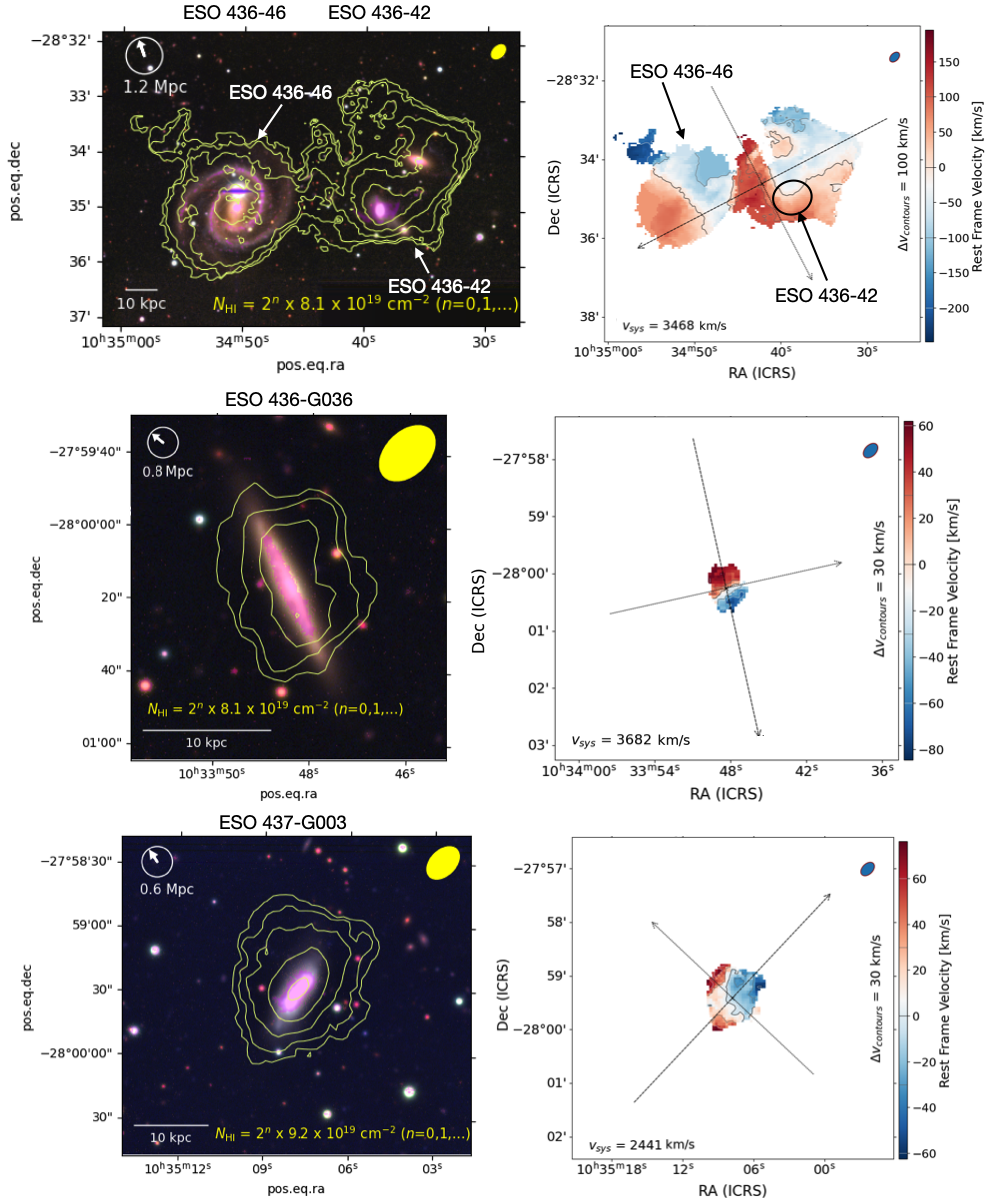}
 \end{subfigure}
 \caption{ Starburst galaxies ESO 436-46 and ESO 436-42, ESO 436- G036, and ESO 437-G003.\\ Left: False rgb color images (red - r band, green - g band, blue -u band) overlaid with and H$\alpha$ map (fuchsia) and with MeerKAT HI contours (yellow). The beam size is displayed at the top right. The compass at the top left points in the direction towards the center of the cluster. The distance indicated below the compass is the projected distance of the galaxy to the center of the cluster. A scalebar of 10kpc can be found at the bottom left.\\ Right: Moment 1 HI-velocity maps.}
 \label{fig:image1}
 \end{figure*}
 \begin{figure*}[hbt!]
 \begin{subfigure}{\columnwidth}
 \includegraphics[width=17.3cm]{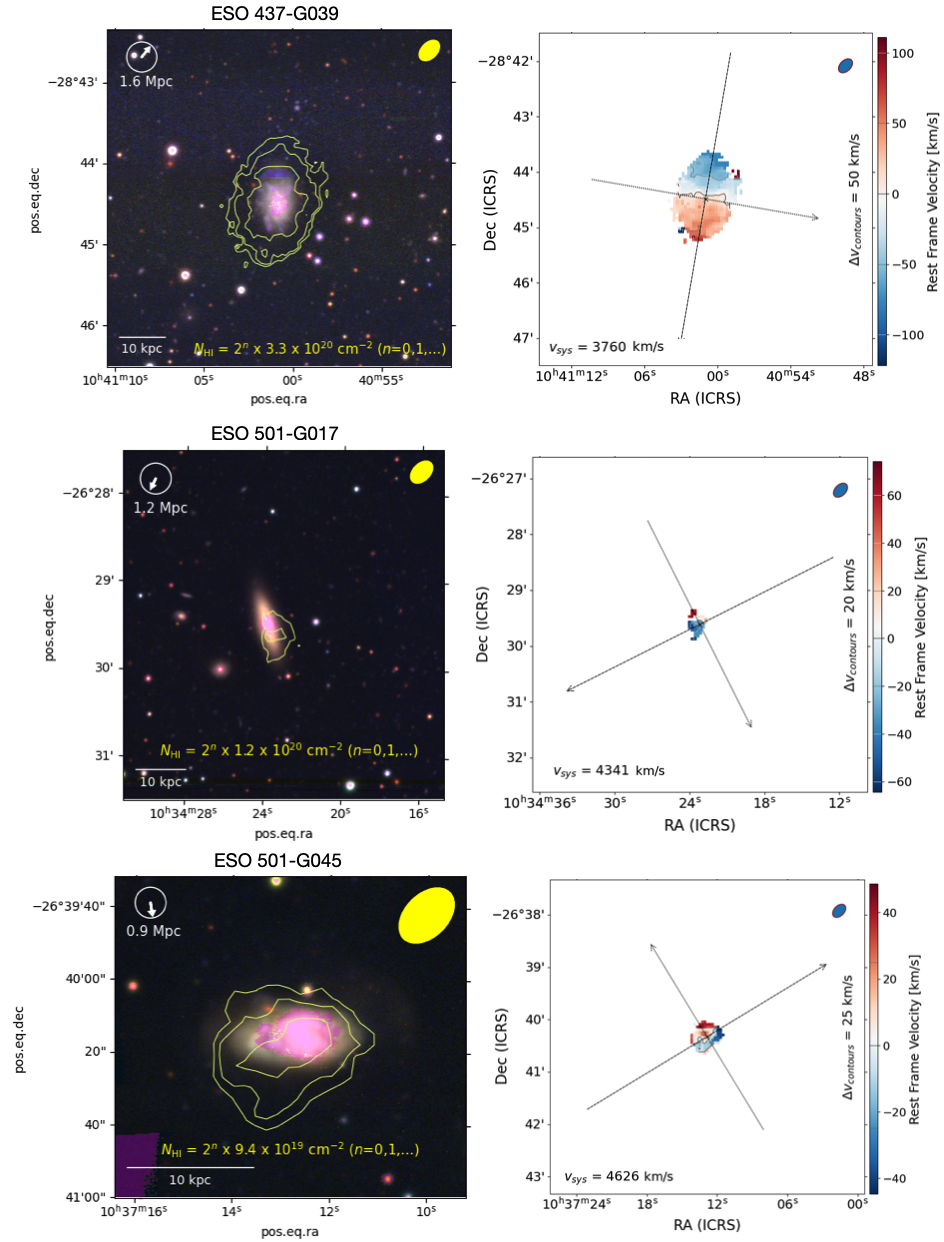}
 \end{subfigure}
 \caption{ Starburst galaxies ESO 437-G039, ESO 501-G017, and ESO 501-G045.\\ Left: False rgb color images (red - r band, green - g band, blue -u band) overlaid with and H$\alpha$ map (fuchsia) and with MeerKAT HI contours (yellow).  The beam size is displayed at the top right. The compass at the top left points in the direction towards the center of the cluster. The distance indicated below the compass is the projected distance of the galaxy to the center of the cluster. A scalebar of 10kpc can be found at the bottom left.\\ Right: Moment 1 HI-velocity maps.}
 \label{fig:image2}
 \end{figure*}

 
 \begin{figure*}[h!]
 \begin{subfigure}{\columnwidth}
 \includegraphics[width=18cm]{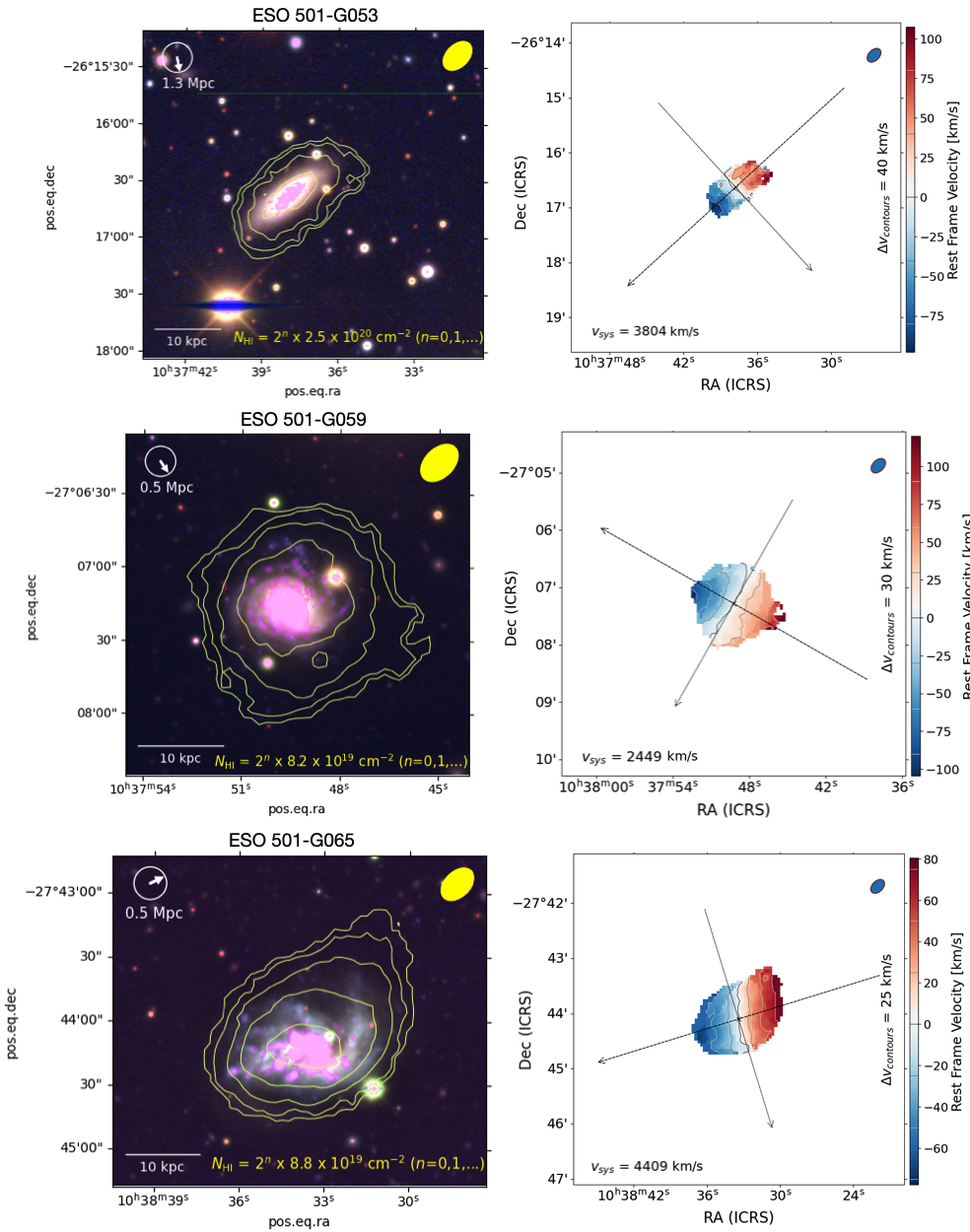}
 \end{subfigure}
 \caption{ Starburst galaxies ESO 501-G053, ESO 501-G059, and ESO 501-G065.\\ Left: False rgb color images (red - r band, green - g band, blue -u band) overlaid with and H$\alpha$ map (fuchsia) and with MeerKAT HI contours (yellow).  The beam size is displayed at the top right. The compass at the top left points in the direction towards the center of the cluster. The distance indicated below the compass is the projected distance of the galaxy to the center of the cluster. A scalebar of 10kpc can be found at the bottom left.\\ Right: Moment 1 HI-velocity maps.}
 \label{fig:image3}
 \end{figure*}

 
 \begin{figure*}[h!]
 \begin{subfigure}{\columnwidth}
 \includegraphics[width=17cm]{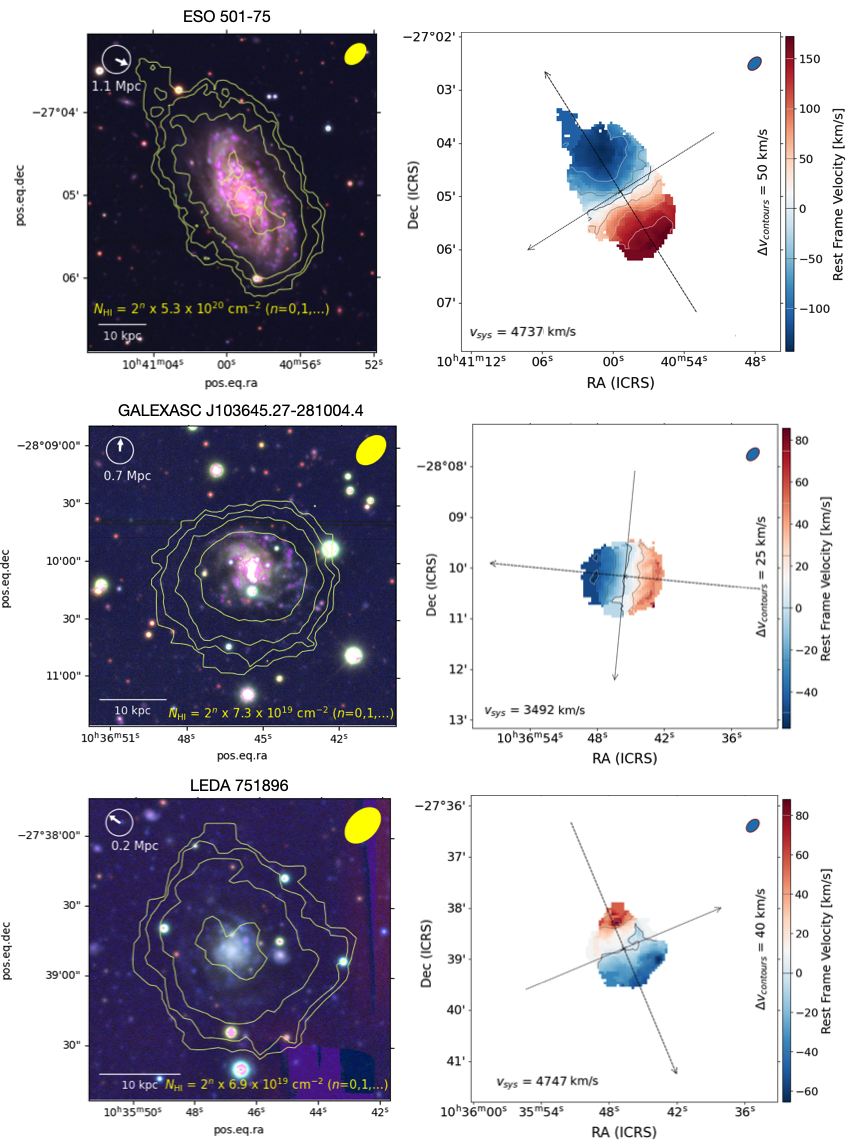}
 \end{subfigure}
 \caption{ Starburst galaxies ESO 501-75, GALEXASC J103645.27-281004.4, and LEDA 751896.\\ Left: False rgb color images (red - r band, green - g band, blue -u band) overlaid with and H$\alpha$ map (fuchsia) and with MeerKAT HI contours (yellow).\\ Right: Moment 1 HI-velocity maps.}
 \label{fig:image4}
 \end{figure*}

 
 \begin{figure*}[h!]
 \begin{subfigure}{\columnwidth}
 \includegraphics[width=16cm]{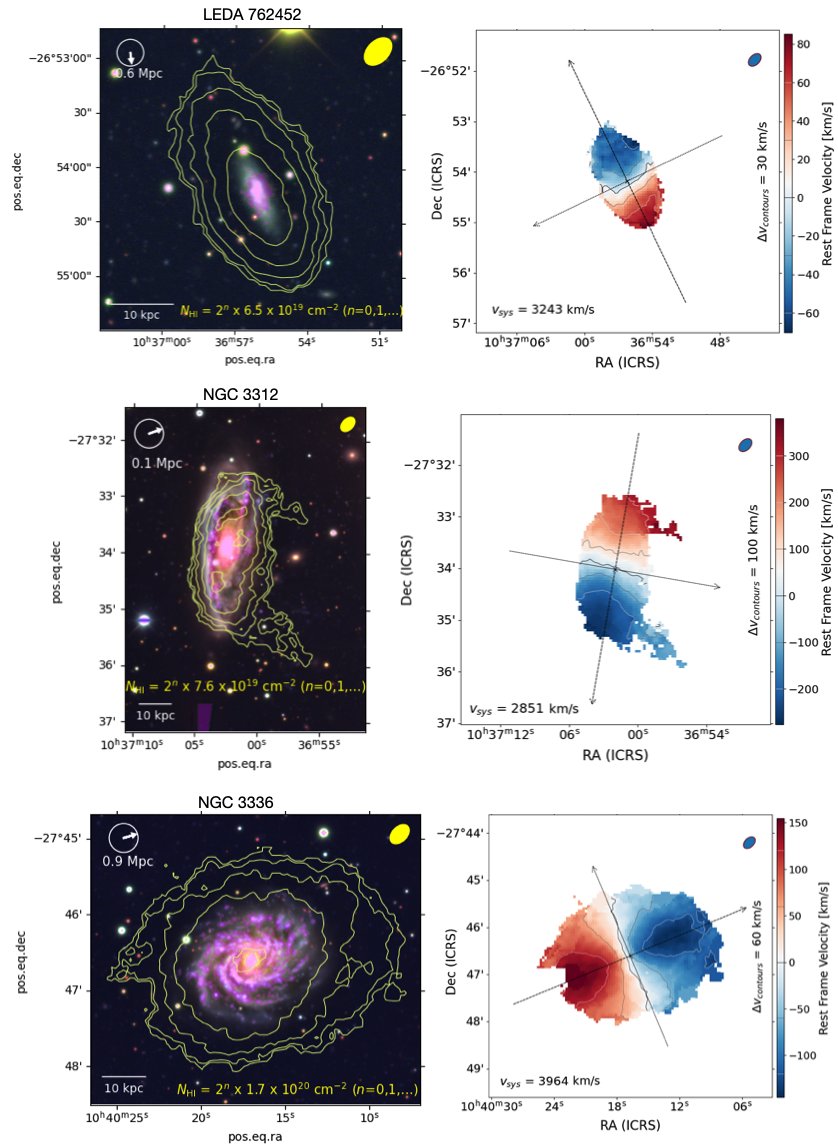}
 \end{subfigure}
 \caption{ Starburst galaxies LEDA 762452, NGC 3312, and NGC 3336.\\ Left: False rgb color images (red - r band, green - g band, blue -u band) overlaid with and H$\alpha$ map (fuchsia) and with MeerKAT HI contours (yellow).  The beam size is displayed at the top right. The compass at the top left points in the direction towards the center of the cluster. The distance indicated below the compass is the projected distance of the galaxy to the center of the cluster. A scalebar of 10kpc can be found at the bottom left.\\ Right: Moment 1 HI-velocity maps.}
 \label{fig:image5}
 \end{figure*}

 
 \begin{figure*}[h!]
 \begin{subfigure}{\columnwidth}
 \includegraphics[width=16cm]{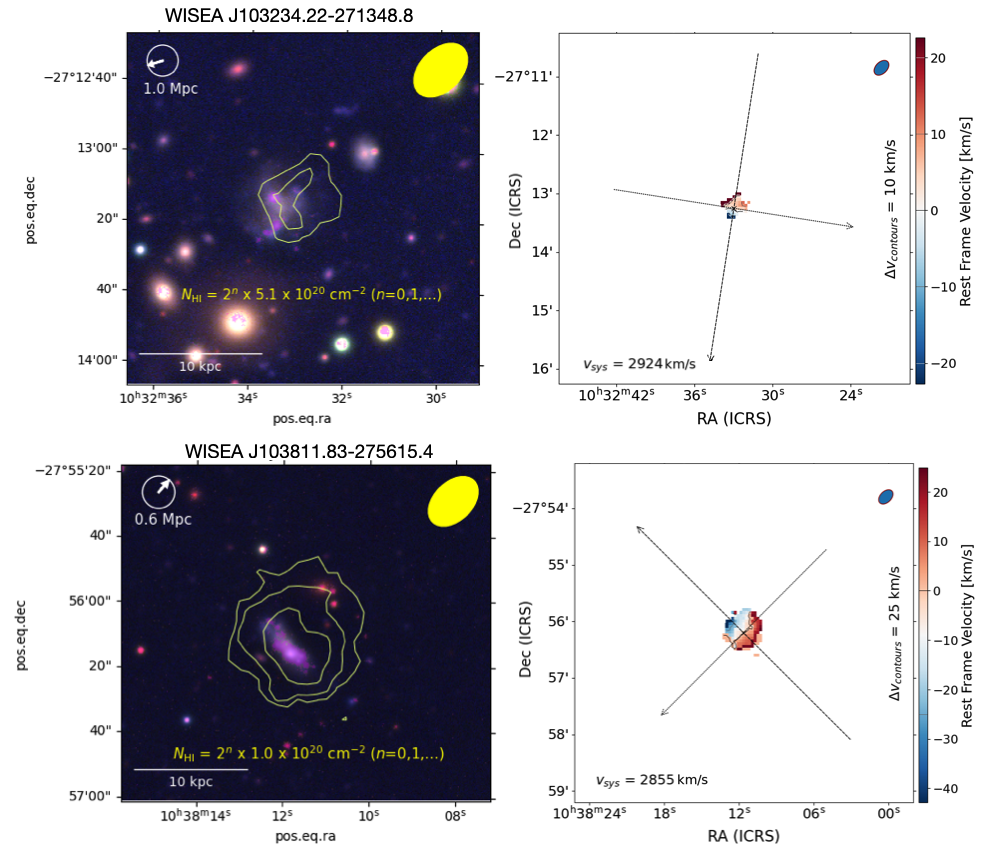}
 \end{subfigure}
 \caption{ Starburst galaxies WISEA J103234.22-271348.8, and WISEA J103811.83-275615.4.\\ Left: False rgb color images (red - r band, green - g band, blue -u band) overlaid with and H$\alpha$ map (fuchsia) and with MeerKAT HI contours (yellow).  The beam size is displayed at the top right. The compass at the top left points in the direction towards the center of the cluster. The distance indicated below the compass is the projected distance of the galaxy to the center of the cluster. A scalebar of 10kpc can be found at the bottom left.\\ Right: Moment 1 HI-velocity maps.}
 \label{fig:image6}
 \end{figure*}

\end{appendix}

\end{document}